\documentclass[apj]{emulateapj}
\usepackage{amsmath}
\usepackage{color}
\usepackage{epsf}
\usepackage{ulem}
\usepackage{graphicx}
\usepackage{natbib}
\bibpunct{(}{)}{;}{a}{}{,}
\RequirePackage{lineno} 
\newcommand{\Om}{\Omega_m}
\newcommand{\Ob}{\Omega_b}

\newcommand{\OL}{\Omega_\Lambda}

\newcommand{\hMpc}{h^{-1}{\rm\;Mpc}}
\newcommand{\hGpc}{h^{-1}{\rm\;Gpc}}
\newcommand{\triGpch}{h^{-3}{\rm\;Gpc^3}}
\newcommand{\itriMpch}{h^{3}{\rm\;Mpc^{-3}}}
\newcommand{\ihMpc}{h{\rm\;Mpc^{-1}}}

\newcommand{\Nb}{{\it N}-body}
\newcommand{\Plin}{P_{\rm lin}}

\newcommand{\Pnwl}{P_{\rm nwl}}
\newcommand{\Pobs}{P_{\rm obs}}
\newcommand{\Zel}{Zel'dovich}
\begin{document}
%\linenumbers
\title{Galaxy Bias and its Effects on the Baryon Acoustic Oscillations Measurements}
\author{
Kushal T. Mehta\altaffilmark{1},
Hee-Jong Seo\altaffilmark{2,3},
Jonathan Eckel\altaffilmark{1},
Daniel J. Eisenstein\altaffilmark{1,4},
Marc Metchnik\altaffilmark{1},
Philip Pinto\altaffilmark{1},
Xiaoying Xu\altaffilmark{1}}
%%%%%%%%%%%%%%%%%%%%%%%%%%%%%%%%%%%%%%%%%%%%%%%%%
\begin{abstract}
The baryon acoustic oscillation~(BAO) feature in the clustering of matter in the universe serves as a robust standard ruler and hence can be used to map the expansion history of the universe. We use high force resolution simulations to analyze the effects of galaxy bias on the measurements of the BAO signal. We apply a variety of Halo Occupation Distributions (HODs) and produce biased mass tracers to mimic different galaxy populations. We investigate whether galaxy bias changes the non-linear shifts on the acoustic scale relative to the underlying dark matter distribution presented by~\cite{Seo09}. For the less biased HOD models ($b < 3$), we do not detect any shift in the acoustic scale relative to the no-bias case, typically $0.10\% \pm 0.10\%$. However, the most biased HOD models ($b > 3$) show a shift at moderate significance ($0.79\% \pm 0.31\%$ for the most extreme case). We test the one-step reconstruction technique introduced by~\cite{ESSS07} in the case of realistic galaxy bias and shot noise. The reconstruction scheme increases the correlation between the initial and final ($z=1$) density fields achieving an equivalent level of correlation at nearly twice the wavenumber after reconstruction. Reconstruction reduces the shifts and errors on the shifts. We find that after reconstruction the shifts from the galaxy cases and the dark matter case are consistent with each other and with no shift.  The $1\sigma$ systematic errors on the distance measurements inferred from our BAO measurements with various HODs after reconstruction are about $0.07\% - 0.15\%$. 
\end{abstract}
%%%%%%%%%%%%%%%%%%%%%%%%%%%%%%%%%%%%%%%%%%%%%%%%%%
\altaffiltext{1}{Steward Observatory, University of Arizona,
933 N. Cherry Ave., Tucson, AZ 85121; kmehta@email.arizona.edu}
\altaffiltext{2}{Berkeley Center for Cosmological Physics, LBL and Department of Physics, University of California, Berkeley, CA, USA 94720}
\altaffiltext{3}{Center for Particle Astrophysics, Fermi National Accelerator Laboratory, P.O. Box 500, Batavia, IL 60510-5011}
\altaffiltext{4}{Center for Astrophysics, Harvard University,
60 Garden St., Cambridge, MA 02138}

%%%%%%%%%%%%%%%%%%%%%%%%%%%%%%%%%%%%%%%%%

\section{Introduction}\label{sec:intro}
Baryon acoustic oscillations~(BAOs) are the impressions of acoustic waves in the hot plasma of the early universe onto the matter distribution. The acoustic waves were imprinted onto the large-scale baryon distribution once the photons and baryons decoupled after recombination with a characteristic wavelength of about 150 Mpc. This characteristic or acoustic scale is the distance traveled by the sound waves before recombination and is known as the ``sound horizon scale'', which has also been measured in the Cosmic Microwave Background~(CMB) to be about one degree. Since galaxies form in overdense regions of the universe, they are preferentially separated by this acoustic scale. The acoustic scale can therefore be used as an excellent standard ruler. More recently, the BAO scale has been measured in large scale surveys as a peak in the correlation function $\xi(r)$ or as a damped harmonic series in the power spectrum $P(k)$~\citep{Eisen05, Cole05, Hutsi06, Tegmark06, Percival07, Blake07, Padmanabhan07, Okumura08, Gaz08, Gaz09, Estrada09, Sanchez09, Percival10, Kazin10}. Due to its robust nature, there has been growing interest in using the BAOs as a powerful probe of dark energy. BAO measurements have become a key component of many dark energy surveys, both current and future. 

The systematic errors on measurements of the acoustic scale are dominated by non-linear structure growth at small scales and low redshift and by redshift-space distortions. Non-linear effects broaden and shift the peak in the correlation function. Hence, it is harder to centroid the peak and obtain precise and accurate measurements of the location of the peak. Redshift distortions further dampen the BAO signal along the line of sight. The largest dark energy surveys plan to measure the acoustic scale to the $0.1\% - 0.2\%$ level. While most of the growth on BAO scales ($\approx$ 150 Mpc) is still linear, to achieve percent and sub-percent level accuracy, it becomes imperative to understand and model the shift and systematic errors caused by non-linear effects. There has been an increasing effort in understanding the effects of non-linear growth and non-linear gravitational flows in order to model them for surveys. Most of the modeling has been done using dark matter simulations that do not incorporate the effects of observing galaxies as mass tracers~\citep{Meiksin99, SE05, ESSS07, Springel06, SSEW08, Angulo08, Nishi09, Taka09, Seo09}. \cite{Angulo08} used semi-analytic models to populate their dark matter halos with galaxy populations (also see \cite{Sanchez08}). However, their high resolution simulation covered a volume of $2.41 \triGpch$. The precision of BAO measurements is proportional to the volume used. In order to simulate real surveys we want to use several tens of $\triGpch$.  In our simulations, we use a variety of galaxy prescriptions but with a volume almost 20 times larger simulation, and explore the effects of reconstruction on the BAO signal.

Redshift surveys observe galaxies and not the underlying matter distribution. It is well known that galaxies form in the high density regions of dark matter halos. Thereby using galaxies as mass tracers introduces an inherent bias into any survey by over-weighting the overdense regions and under-weighting the underdense regions. \cite{PW09} use perturbation theory to show that we should expect the shift on the acoustic scale to be less than a percent even with the presence of galaxy bias. In order to measure this effect precisely,  we need to run simulations with the effects of galaxy bias. While \cite{SSEW08, Seo09} show the shift in the acoustic scale due to nonlinear structure growth and redshift distortions, here we include the effect of galaxy bias in our highest resolution simulation set. In this work, we use a set of high resolution simulations with various Halo Occupation Distributions~(HODs) to model the galaxy populations expected to be measured by various surveys~\citep{Zheng07, Zheng09}. We investigate the effects of galaxy bias on the acoustic scale for each case. We compare our results with galaxy bias to the results from~\cite{Seo09}, who used all dark matter particles to trace the matter density field (hence, no galaxy bias). 

\cite{ESSS07} presents a simple, one-step reconstruction technique to account for the effects of large-scale gravitational bulk flows that degrade the acoustic signal. This method has been proved to be successful for dark-matter simulations even for non-negligible shot noise levels~\citep{ESSS07, SSEW08, Seo09, Noh09}. \cite{Huff07} look at the effect of galaxy bias on reconstruction on a smaller volume (1.1 $\triGpch$). In this paper, we use a much larger volume of 44 $\triGpch$ to investigate the effect of galaxy bias on the technique of reconstruction.

In \S~\ref{sec:sims} we describe the simulations, the Halo Occupation Distributions, fitting methods and the reconstruction technique. We introduce the propagator and analyze the effects of galaxy bias and reconstruction on it in \S~\ref{sec:Ck}. Next, we compute the acoustic scale and investigate the effects of galaxy bias on the acoustic scale in \S~\ref{sec:alpha}. We also scompare our results to the Fisher Matrix estimates using the code provided by \cite{SE07} and to the dark matter only analysis of \cite{Seo09}. We summarize our results in \S~\ref{sec:conc}.

\section{Simulations \& Methods}\label{sec:sims}
\subsection{Simulations and Code}
We generate high force resolution \Nb~simulations using the ABACUS code by~\cite{MM10}. The ABACUS code uses a new method to compute \Nb~forces for periodic boundary conditions by computing the force on a particle from the rest of the simulation and an infinite sum of periodic images. The force on a particle is calculated by dividing the simulation into near field and far field. The near field force is computed by a direct $O(N^2)$ method with a gravitational softening length of $0.0488 \hMpc$. The far-field force is computed by summing the forces resulting from the multipole expansion of the mass distribution within a grid of distant cells, aided by a cyclic convolution to speed the calculation.

\begin{figure}
\centering
\includegraphics[width=6.5in]{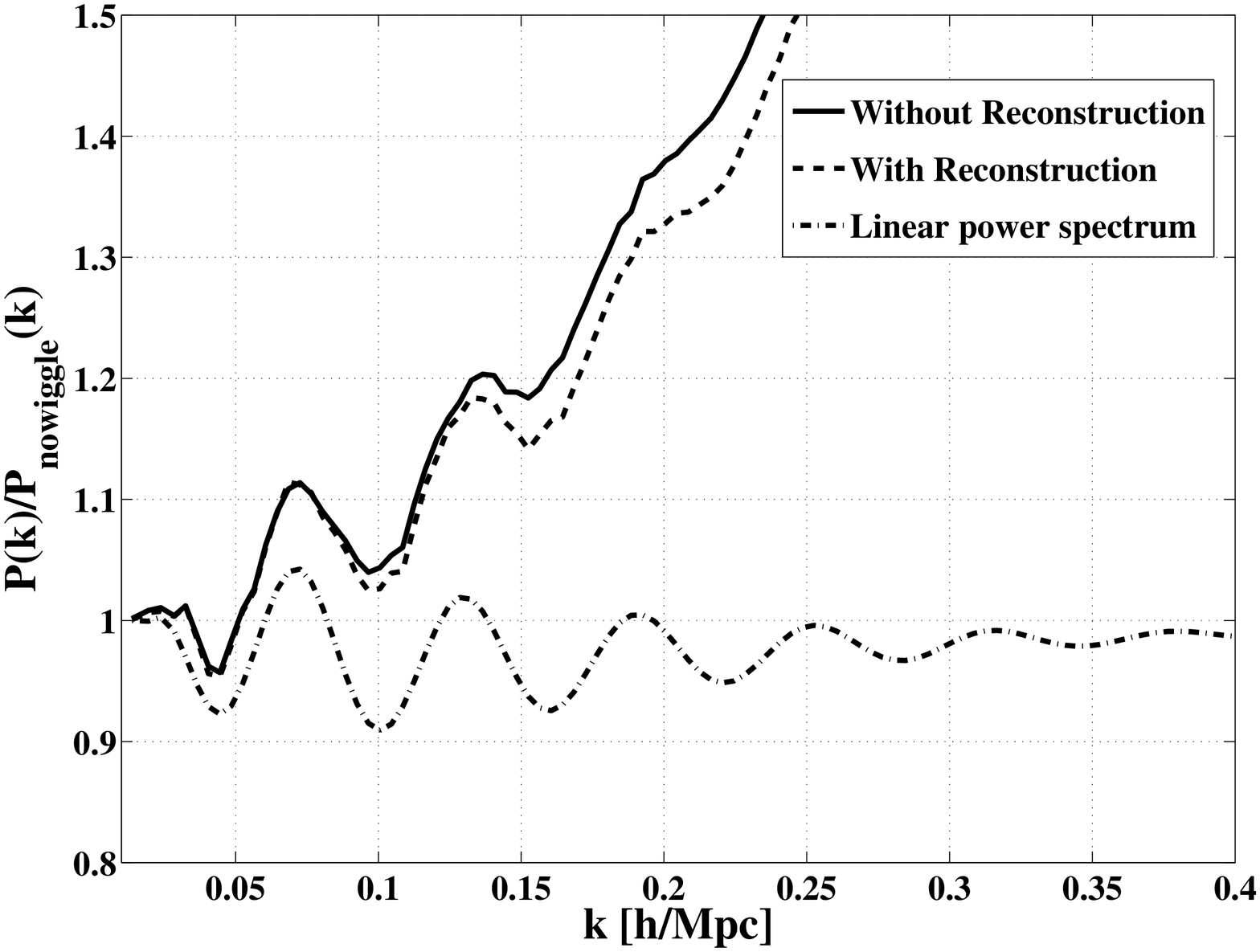}
\caption{The average power spectrum from 44 simulations (HOD 1a, real space) divided by the no wiggle power spectrum, $P(k)/Pnwl(k)$. The no-wiggle power spectrum represents the power spectrum without the BAO peaks~\cite{EH98}. Thus, we clearly see the BAO peaks
appear in the power spectrum. We also start to see the increasing power on small scales (large $k$) as non-linear growth starts to dominate. Plotted in the dashed line is the average power spectrum after reconstruction. We see that our reconstruction scheme reduces the effect of non-linear structure formation on small scales and restores information into the acoustic peaks. The dot-dashed line representes the linear power spectrum divided by the no-wiggle power spectrum. }
\label{fig:BAOpeaks}
\end{figure}

 We assume concordance cosmology consistent with the WMAP5+SN+BAO results~\citep{Komatsu09}: $\Om = 0.279$, $\OL = 0.721$, $h = 0.701$, $\Ob = 0.0462$, $n_s = 0.96$, $\sigma_8 = 0.817$. We used $2^{\rm nd}$ order Lagrangian perturbation theory~(2LPT) based IC code by \cite{Sirko05} to generate the initial conditions. In particular, we use linear theory density field at $z = \infty$  and use 2LPT code to generate the density field at $z = 50$. The initial ($z = \infty$) density field is computed by back-evolving the $z = 0$ output from CMBFAST assuming a matter dominated universe. When we compare our results to the initial density field, we always use the initial linear theory density field. We evolve the density fields from $z = 50$ to $z = 1$ using the ABACUS code and calculate the spherically averaged power spectrum in both real space and redshift space. We generate 44 simulations each containing $1024^3$ particles in a $1 \hGpc$ box giving us a total volume of $44 \triGpch$. The resolution of our simulations gives us a particle mass of $7.2 \times 10^{10} h^{-1} M_\odot$. Note that this is the same simulation set as G1024 set presented in \cite{Seo09}.  Figure~\ref{fig:BAOpeaks} shows the effects of structure evolution on the baryon acoustic oscillations in the power spectrum at $z = 1$. In this figure, the power spectrum is divided by the no-wiggle power spectrum $\Pnwl (k)$, which is the matter density power spectrum without the BAO harmonic features~\citep{EH98}. We see two distinct harmonic acoustic peaks before the acoustic oscillations get erased by non-linear structure growth on small scales. 

\subsection{Halo Occupation Distributions (HODs)}\label{sec:HOD}
Perturbation theory explains how overdensities in matter grow while dark matter simulations help us better understand non-linear effects. In real surveys, of course, we do not observe the dark matter particles but galaxies residing in dark matter halos. We assume that the galaxy population are tracers of the underlying dark matter distribution. Since galaxies form in the highest density regions of dark matter halos, they represent a biased version of the matter density. It is critical that we understand how biased tracers of the matter distribution affects the acoustic scale. To examine the effects of galaxy populations, we generated various Halo Occupation Distributions (HODs) and applied them to our simulations~\citep{Peacock00, Berlind02}. The HOD models allow us to use various mass tracers (galaxy populations) to study the acoustic scale. To generate the HODs, we use a simple Friends-of-Friends (FoF) program \citep{Davis85} to identify collapsed dark matter halos with a linking length of $b_{FoF} = 0.16$. We populate each halo with a central and satellite galaxies using the following formula:
\begin{equation}
N_{g}(M) = 
\begin{cases}
0 & \text{if $M < M_{cen}$} \\
1 + \text{Poisson}(M/M_{sat}) & \text{if $M > M_{cen}$}
\end{cases}
\label{eq:HOD}
\end{equation}
where $M$ is the halo mass, $M_{cen}$ is the minimum mass needed to have a central galaxy, Poisson$(M/M_{sat})$ is an integer randomly chosen from a Poisson distribution with mean $M/M_{sat}$, and $M_{sat}$ is the minimum mass to have at least one satellite galaxy \citep{Guzik02, Berlind03, Kravtsov04, Zheng05}. Note that $P(M/M_{sat}) = 0$ for $M < M_{sat}$. To test the effects of various galaxy populations, we vary $M_{cen}$ and $M_{sat}$ to generate 12 different HOD models whose properties and definitions are detailed in Table~\ref{tab:HOD}. By varying the mass thresholds $M_{cen}$ and $M_{sat}$, we obtain HOD models with different number of galaxies. For each central galaxy threshold, $M_{cen}$, we vary the satellite mass threshold and then slightly adjust $M_{cen}$ to give us the total number of galaxies: $2 \times 10^6$, $1 \times 10^6$, $3 \times 10^5$ and $1 \times 10^5$. The central galaxy is assigned the halo's center of mass position and velocity. We then randomly pick a corresponding number of halo particles and assign their positions and velocities to the satellites.

Thus, each HOD identifies halos based on the mass threshold and replaces the halo by a central galaxy and satellite galaxies (based on the satellite threshold) and uses these galaxies for any analysis. We use the various HOD models as biased tracers of the density fields and explore their effects on the acoustic scale. Since we seek to understand the effects of various galaxy populations or HOD models, we use the ``mass'' case, where we do not apply any HOD, to be our base case. A full analysis of the mass case is presented in~\cite{Seo09}. The different number densities corresponding to each HOD also allow us to explore the effect of shot noise on the acoustic scale. 

\subsection{Power Spectra Fitting Methods}\label{sec:fits}
We compute the power spectrum for each of the 44 simulations following the method used by \cite{SSEW08, Seo09}. We use $\chi^2$ analysis to fit the spherically averaged power spectrum, $\Pobs$, to the template power spectrum $P_m(k)$:
\begin{equation}
\Pobs = B(k)P_m(k/\alpha) + A(k),
\label{eq:Pobs}
\end{equation}
where $B(k)$ represents scale-dependent bias and redshift distortions; $A(k)$ represents anomalous power such as shot noise and additive terms of non-linear growth; and $\alpha$ represents the ratio of the linear acoustic scale to the measured acoustic scale. Therefore, $\alpha~-~1$ measures the shift in the acoustic scale i.e. the error in the distance inferred if one used Eq.~\eqref{eq:Poly7} as the template. $\alpha~>~1$ implies that the measured acoustic scale is shifted towards smaller scales or larger Fourier wavenumbers. In this paper, we quote our shifts in the acoustic scale as $\alpha~-~1~(\%)$ in percent. Our template power spectrum $P_m(k)$ is constructed by modifying the linear power spectrum. We use a nonlinear parameter $\Sigma_M$ that degrades the BAO peaks in the linear power spectrum to account for nonlinear effects and redshift distortions. The form of the template power spectrum is given by:
\begin{equation}
P_m(k) = [\Plin(k) - \Pnwl(k)]\exp\left(\frac{-k^2\Sigma_M^2}{2}\right) + \Pnwl(k)
\label{eq:Pm}
\end{equation}
where $\Plin (k)$ is the linear power spectrum and $\Pnwl (k)$ is the no-wiggle power spectrum as described by~\cite{EH98}. Due to the large degree of polynomials used to fit $A(k)$ and $B(k)$~(Eq.~\eqref{eq:Poly7}), our results are not sensitive to our value of $\Sigma_M$. We use the fiducial values for $\Sigma_M$ given by the \Zel~approximation: $\Sigma_M ~ 5.3 \ihMpc$ for real space and $~7.0 \ihMpc$ for redshift space. As shown by \cite{SSEW08}, the fitting results are not sensetive over a range of $\Sigma_M$ values ($\Delta \Sigma_M = \pm 2 \ihMpc$) due to the large degree of polynomials used to fit $A(k)$ and $B(k)$. We use a $7^{th}$ order polynomial for $A(k)$ and a $2^{nd}$ order polynomial for $B(k)$ for both real space and redshift space and fit over $40.02 \ihMpc < k < 0.35\ihMpc$:
\begin{center}
\begin{eqnarray}
A(k) & = & a_0 + a_1k + a_2k^2 + ... + a_7k^7 \text{~~and} \nonumber \\
B(k) & = & b_0 + b_1k + b_2k^2.
\label{eq:Poly7}
\end{eqnarray}
\end{center}

\subsection{Resampling Simulations Method}\label{sec:Resampling}
In order to measure the scatter in $\alpha$ and therefore the scatter in the acoustic scale, we use a modified bootstrap method. We generate 1000 subsamples by randomly selecting M boxes without replacement out of a total of N simulations. We then perform a $\chi^2$ fitting to get the best fit $\alpha$ for the individual subsamples and find the mean $\alpha$ and the scatter. The scatter in the measured $\alpha$'s is rescaled by a factor of $\sqrt{M/(N-M)}$ to reflect the scatter in the mean $\alpha$. For our resampling method, we have N = 44 and we choose M = 22 so that the rescaling factor is unity. To analyze the scatter in $\alpha$, we fit each simulation assuming a covariance matrix of the power spectrum given by a Gaussian random field. This method reflects any non-Gaussianity in the density field in the scatter in the best-fit $\alpha$'s between different simulations.

\subsection{Reconstruction with HODs}\label{sec:Recon}
Large-scale gravitational forces cause large scale velocity fields between the overdensities. These velocity fields tend to broaden the acoustic peak and degrade the BAO measurement. We employ a simple reconstruction scheme introduced by~\cite{ESSS07}. This reconstruction scheme is based on the \Zel~approximation and models the large scale velocity fields to attempt to restore the BAO information and restore information about the initial linear density field.

In linear perturbation theory, we can estimate the displacements of mass particles based on the density perturbations using the \Zel~approximation. With reconstruction, we estimate the bulk flows based on the large-scale information of the observed nonlinear density field and undo the large scale velocity fields to recover the portion of BAO that has been degraded. The Lagrangian particle position $\vec{r}$ can be mapped to an Eulerian particle position $\vec{x}$ by the displacement $\vec{q}$:
\begin{equation}
\vec{x} = \vec{r} + \vec{q}.
\label{eq:Lag}
\end{equation}
The density in Eulerian coordinates is transformed to the density in Lagrangian coordinates through the Jacobian:
\begin{center}
\begin{eqnarray}
\rho_0 \; d\vec{r} & = & \rho \; d\vec{x} \\
\frac{\rho}{\rho_0} & = & 1 + \delta = \frac{d\vec{r}}{d\vec{x}} = J^{-1} = \frac{1}{\text{Det}(\delta_{ij}^K + \partial q_i/\partial r_j)}
\label{eq:rho}
\end{eqnarray}
\end{center}
where $\rho_0$ is the Lagrangian density and is the same as the Eulerian mean density, $\delta$ is the Eulerian overdensity, and $\delta^K$ is the Kronecker delta function. If we expand the expression to linear order, it becomes:
\begin{equation}
\delta = -\nabla\cdot\vec{q}.
\label{eq:cont}
\end{equation}
In the linear regime growing mode in CDM cosmologies, $\vec{q}$ is a curl-free field. Hence, we can define a scalar field $\phi$ such that:
\begin{equation}
\vec{q} = -\nabla\phi.
\label{eq:phiintro}
\end{equation}
In our reconstruction scheme, we are looking to undo large scale velocity flows and thus we smooth the small-scale perturbations by using a Gaussian filter. Thus, we enter only the large-scale information to derive the bulk flows. In our analysis, we use a Gaussian smoothing width of $R = 14 \hMpc$ to obtain the real-space galaxy density field $\delta^{\rm real}$ in configuration space. Using Eq~\eqref{eq:cont} and Eq~\eqref{eq:phiintro}, we can express $\phi$ in terms of $\delta^{\rm real}$ as follows:
\begin{center}
\begin{eqnarray}
\nabla^2 \phi^{\rm real} & = & \frac{\delta^{\rm real}}{b} \text{~~and} \\
\phi^{\rm real} & = & -\int\frac{d^3k}{(2\pi)^3}\frac{\tilde{\delta}^{\rm real}}{b}\frac{1}{k^2}e^{i\vec{k}\cdot\vec{r}}
\label{eq:phireal}
\end{eqnarray}
\end{center}
where $\tilde{\delta}^{\rm real}$ is the real-space smoothed density field in Fourier space. While we use Eq~\eqref{eq:phireal} to solve Eq~\eqref{eq:phiintro} in Fourier space, it can also be easily solved in configuration space as a central force problem. Hence, we call the displacement field derived from solving Eq~\eqref{eq:phiintro} the trivial displacement field $\vec{q}_{\rm ~trivial}^{\rm ~real}$, which is given by:
\begin{eqnarray}
\nabla\cdot\vec{q}_{\rm ~trivial}^{\rm ~real} & = & -\frac{\delta^{\rm real}}{b} \text{~~and therefore} \nonumber \\
\vec{q}_{\rm trivial}^{\rm real} & = & -\nabla\phi^{\rm real}
\label{eq:qtrivreal}
\end{eqnarray}
In real space, this trivial displacement field is the desired displacement field. In our simulations, we displace the data points and the random points by the displacement field $\vec{q}_{\rm ~trivial}^{\rm ~real}$. By displacing both the data and random points, we ensure that the amplitude of the overdensities remains the same while removing the effects of the large scale velocity field. 

In galaxy redshift surveys, the density field we observe is subject to redshift distortions. We can relate the redshift-space density field $\tilde{\delta}^{\rm red}$ in Fourier space to that in real space as follows:
\begin{equation}
\tilde{\delta}^{\rm red} = (1 + \beta\mu^2)\tilde{\delta}^{\rm real},
\label{eq:deltared}
\end{equation}
where $\mu = k_z/\left|k\right|$ and $z$ is the line-of-sight direction. We can then compute the redshift-space scalar field $\phi^{\rm red}$ from Eq~\eqref{eq:phireal} and Eq~\eqref{eq:deltared}:
\begin{equation}
\phi^{\rm red} = -\int\frac{d^3k}{(2\pi)^3}\frac{\tilde{\delta}^{\rm red}}{b}\frac{1}{k^2}\frac{1}{1 + \beta\mu^2}e^{i\vec{k}\cdot\vec{r}}.
\label{eq:phired}
\end{equation}
Note that $\phi^{\rm red}$ is our estimate for the scalar field that generates the real-space displacement. Next we relate $\phi^{\rm red}$ to the redshift-space galaxy density field:
\begin{center}
\begin{eqnarray}
\nabla\cdot\vec{q}_{\rm ~trivial}^{\rm ~red} & = & -\frac{\delta^{\rm red}}{b} = -\left(\nabla^2 + \beta\frac{\partial^2}{\partial z^2}\right)\phi^{\rm red} \\ \nonumber
& = & -\nabla\cdot\left(\nabla\phi^{\rm red} + \beta\hat{z}\frac{\partial\phi^{\rm red}}{\partial z}\right).
\label{eq:qdeltaphi}
\end{eqnarray}
\end{center}
It follows from Eq~\eqref{eq:qdeltaphi}, that $\vec{q}_{\rm ~trivial}^{\rm ~red}$ can be expressed as:
\begin{equation}
\vec{q}^{\rm ~red}_{\rm ~trivial} = -\nabla\phi^{\rm ~red} + \beta\hat{z}\frac{\partial\phi^{\rm red}}{\partial z}.
\label{eq:qtrivred}
\end{equation}
From Eq~\eqref{eq:qtrivred}, we can relate $\vec{q}_{\rm ~trivial}^{\rm ~red}$ to $\vec{q}^{\rm ~real}_{\rm ~trivial}$ as follows:
\begin{center}
\begin{eqnarray}
\vec{q}_{\rm ~trivial,~z}^{\rm ~red} & = & (1 + \beta)\vec{q}_{\rm ~trivial,~z}^{\rm ~real}, \\ \nonumber
\vec{q}_{\rm ~trivial,~x,y}^{\rm ~red} & = & \vec{q}_{\rm ~trivial,~x,y}^{\rm ~real},
\label{qredqreal}
\end{eqnarray}
\end{center}
where $x$ and $y$ are the directions orthogonal to the line-of-sight. Finally, we note that the redshift-space displacement field is distorted along the $z$-direction and hence we must modulate the field by a factor of $1 + f$ in the $z$-direction, where $f = d\ln D/d\ln a$ is the logarithmic derivative of the growth function. The redshift-space displacement field is then given by:
\begin{center}
\begin{eqnarray}
\vec{q}_{~z}^{\rm ~red} & = & \frac{1 + f}{1 + \beta}\vec{q}^{\rm ~red}_{\rm ~trivial; ~z}, \\
\vec{q}_{~x,y}^{\rm ~red} & = & \vec{q}^{\rm ~red}_{\rm ~trivial; ~x,y}.
\label{eq:qred}
\end{eqnarray}
\end{center}
Thus, in redshift space, we displace our data points and random points by the displacement field $\vec{q}^{\rm ~red}$ given by Eq~\eqref{eq:qred}. In later sections, we demonstrate that this simple reconstruction techniques improves the correlation between density fields, reduces the measured shift in the acoustic scale and reduces the scatter around the measured shift. We plan to explore the effects of moving the data and random points by different amounts on reconstruction in future work. 

\section{Investigating the Correlation of Density Fields}\label{sec:Ck}
\subsection{Introduction to the Propagator}\label{sec:Ckintro}
In linear perturbation theory, the Fourier modes grow independently of each other, e.g., there is no mode-coupling. In this section, we test how close we are to linear theory on a mode-by-mode basis. We look at the correlation of initial and final ($z = 1$) density field for each mode via the propagator, which compares the amount of the initial density field retained in each mode of the final density field. We derive the propagator in the presence of galaxy bias and compare it to the propagator for the mass case presented by~\cite{Seo09} to estimate the additional BAO signal erased by galaxy bias. 
\begin{center}
\begin{equation}
G(k) = \frac{1}{b \times P_{initial}(k)} \left\langle\tilde{\delta}_{initial}(\vec{k})~\frac{\tilde{\delta}_{final}^*(\vec{k})D(z=\infty)}{D(z=1)(1 + \beta\mu^2)}\right\rangle
\label{eq:Ck}
\end{equation}
\end{center}
Here $\delta_{initial}$ is the linear theory density field at $z = \infty$, $\delta_{final}$ is the density field at $z = 1$, $D(z)$ is the growth function, $P_{initial}(k)$ is the power spectrum of the initial density field, and $b$ is the clustering bias for a given biased tracer.

\begin{figure}
\centering
\includegraphics[width=6.5in]{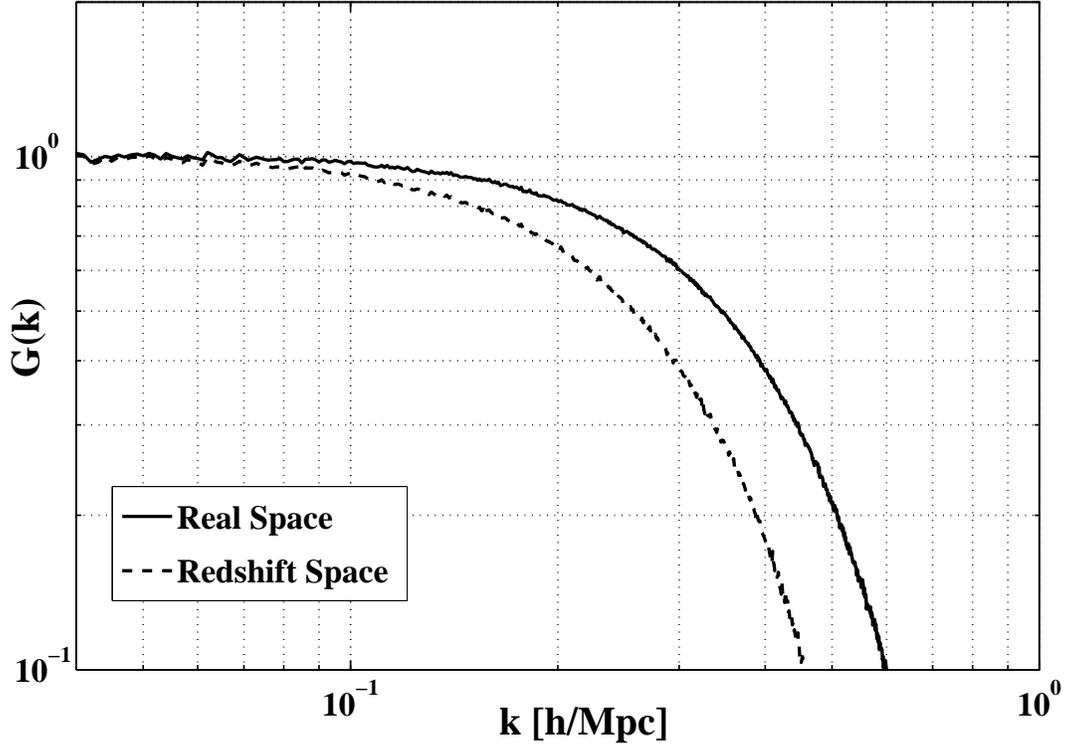}
\caption{The correlation between the initial and final density field as described by Eq.~\eqref{eq:Ck} for HOD 1a. The solid line represents real space while the dashed line represents redshift space. We see that redshift distortions reduce the correlation between the initial and final ($z = 1$) density fields. We are most interested in the correlation between the density fields at BAO wavenumbers ($k \approx 0.09 - 0.2 \ihMpc$). }

\label{fig:Ckrealred}
\end{figure}

Matter growth is linear on large scales and hence we expect that the initial and final density fields should be extremely correlated regardless of the mass tracer used. In the absence of non-linear effects and galaxy bias, we expect perfect correlation between low redshift and high redshift density fields which is represented by $G(k) = 1$. Figure~\ref{fig:Ckrealred} shows the propagator, $G(k)$ for real and redshift space for HOD1a. From Figure~\ref{fig:Ckrealred}, we note that at $k \approx 0.1~\ihMpc$, the wavenumber around which most of the acoustic information is contained, both the real and redshift space density fields are correlated extremely well with the initial density field: $G(k) \geq 0.9$. As expected, the non-linear effects and redshift distortions become increasingly dominant at smaller scales causing the steep drop-off in the propagator. We also see that the real space density field is correlated with the initial density field up to to much smaller scales (larger $k$) than redshift space. The correlation in redshift space is lower on large scales than real space due to redshift distortions.

\subsection{Effect of Galaxy Bias on the Propagator}\label{sec:CkHOD}
We next use the HOD models from \S~\ref{sec:HOD} to evaluate impact of galaxy bias on the recovery of the initial density field. By using the fact that the initial and final densities must be correlated with $G(k) = 1$ at small $k$, we can derive the bias for every HOD. In this paper, we use the real space $G(k)$ to derive the bias. The bias is reflected in the propagator as a multiplicative factor. If we force $b = 1$ in Eq.~\eqref{eq:Ck}, then $G(k)$ will converge to the bias at small wavenumber. We fit a constant function to the real space propagator, $G(k)$, for $k~<~0.08~\ihMpc$ to compute the bias for each HOD.

For each HOD, we want to estimate the amount of noise increase due to bias relevant to the signal. We define an effective number density ($n_{\it eff}$) by Equation~\eqref{eq:neff} and take any deviation from linear bias into account rather than merely using the shot noise. By definition, this also parametrizes the degree of the scale dependence of the bias.
\begin{center}
\begin{eqnarray}
\frac{1}{n_{\it eff}} & = & P_{\rm HOD}(k = 0.2~\ihMpc) \\ \nonumber
& & - {\rm bias}^2 \times P_{\rm lin}(k = 0.2~\ihMpc)
\label{eq:neff}
\end{eqnarray}
\end{center}
where $P_{\rm lin}$ is the linear power spectrum and $P_{\rm HOD}$ is the measured HOD power spectrum. Table~\ref{tab:HOD} gives the definitions, HOD parameters, biases, effective number density and $n_{\it eff}P(k = 0.2 \ihMpc)$ for each HOD. From Table~\ref{tab:HOD}, we see that we span a wide range of effective number densities ($10^{-3} - 10^{-4.1}$) and a range of $n_{\it eff}P$ ($2.6 - 0.6$). 

\begin{figure}
\centering
\includegraphics[width=6.5in]{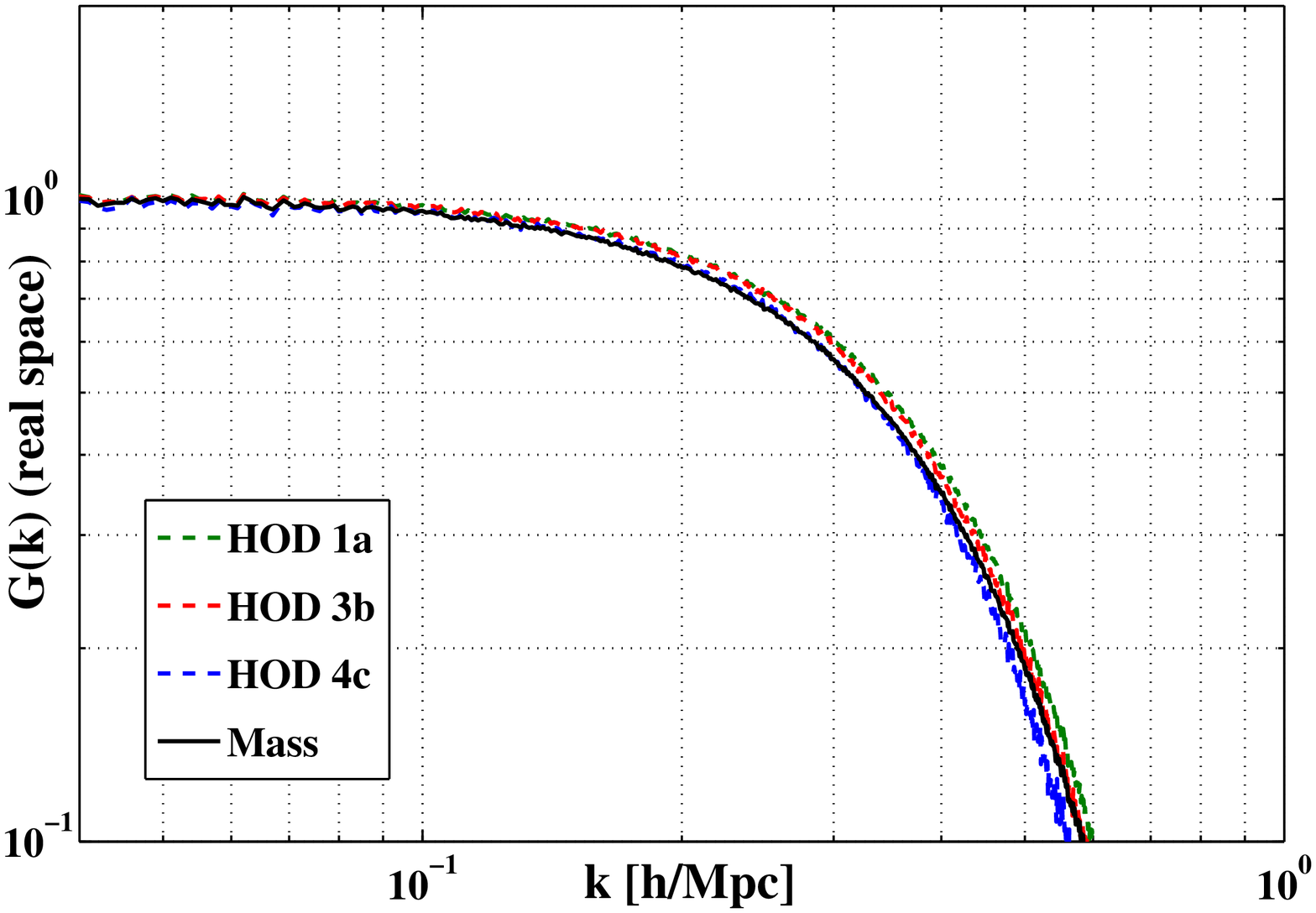}
\includegraphics[width=6.5in]{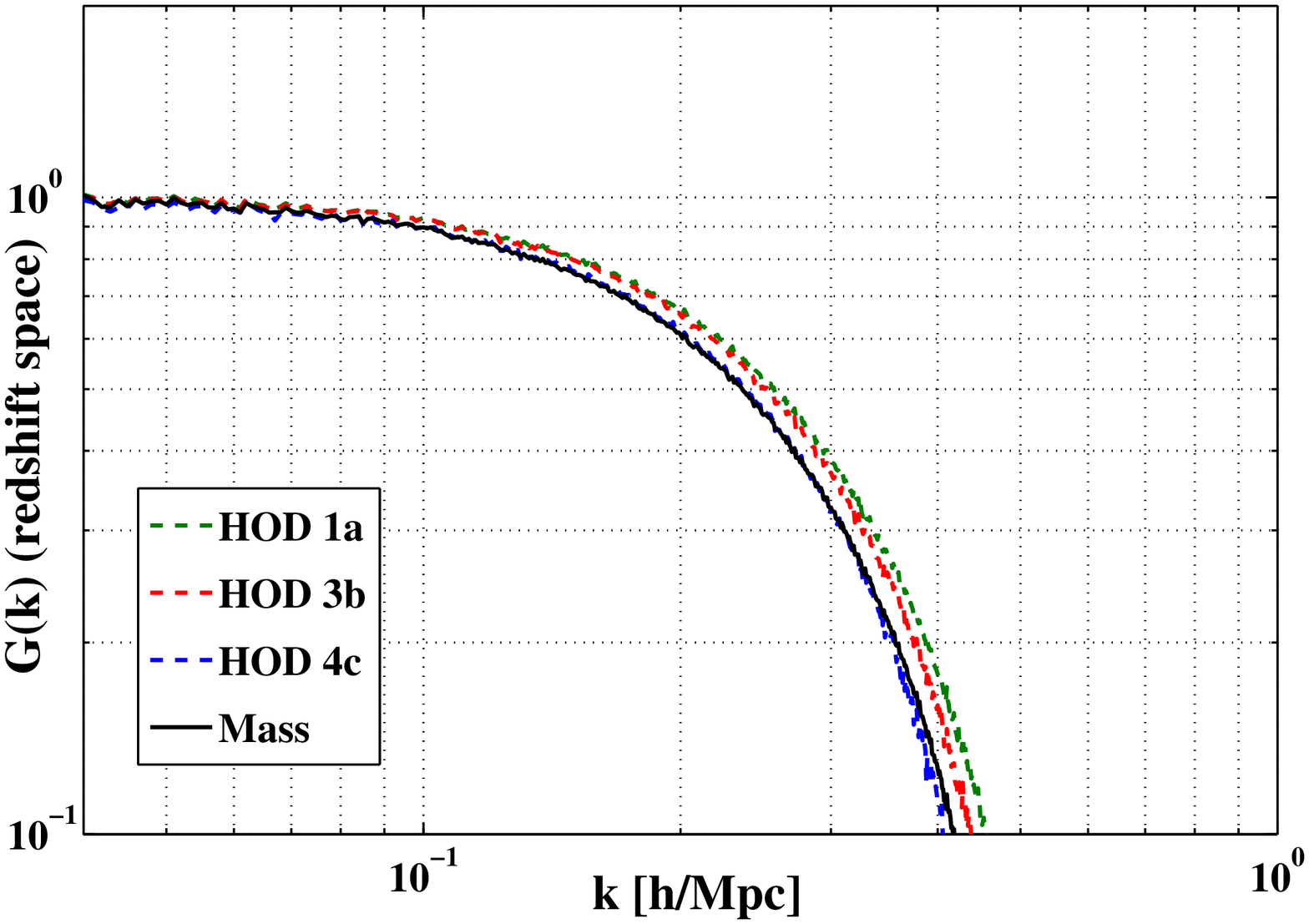}
\caption{The propagator, $G(k)$ for three HOD models and the mass case. The top panel shows real space and the bottom panel shows redshift space. We see that while redshift distortions reduce the correlation between the density fields, they do not spread the individual HOD models from the mass case.}
\label{fig:Ckrs}
\end{figure}

Since we are looking at effects of various HOD models, we calibrate our results against the mass case (no HOD), which along with three of the twelve HOD models are shown in figure~\ref{fig:Ckrs} for real space (upper panel) and redshift space (lower panel). The three HOD models shown in figure~\ref{fig:Ckrs} represent the spread in the propagators for all HOD models and the other HOD models give very similar results. In real space, we see the $G(k)$ plots for various HOD models are very similar with each other. The propagators for different HOD models diverge from each other at large $k$. We attribute this to shot noise which depends on the number density of each HOD. Hence, the effect of bias on the propagator in real space is very small to the acoustic wavenumbers. In redshift space, the separation between the HOD models is more pronounced because of the Finger of God~(FoG) distortions in redshift space. Since different HOD models pick out halos at different mass thresholds, the FoG distortions are different for each HOD model. However, we note that in both real and redshift space, the separation between the HOD models at the acoustic wavenumbers is small.

Figure~\ref{fig:Ckrs} also shows the important result that the propagators for the HOD models are very similar to the mass case. In fact, we see that the biased tracers are slightly better than the mass case. This is consistent with a similar result found by \cite{Noh09}. We can explain this effect as follows. The biased cases replace the dark matter halo with a central galaxy and fewer and satellite galaxies. The central galaxies have the same proper velocity as the center of mass of their halo while the satellite galaxies have the proper velocity of the corresponding dark matter particles. Therefore, the biased cases have experienced less random motion than the mass case. Hence, the halos in the mass case are more extended than in the biased case. In the mass case, we smear out the density field and erase some of the correlation with the initial field compared to the biased cases. We see that this effect is amplified in redshift space due to redshift distortions. While the propagator results are consistent with our explaination, we do not rule out effects of halo clustering as explored by \cite{Angulo08}.

\subsection{Effect of Reconstruction on the Propagator}\label{sec:CkRecon}
\begin{figure}
\centering
\includegraphics[width=6.5in]{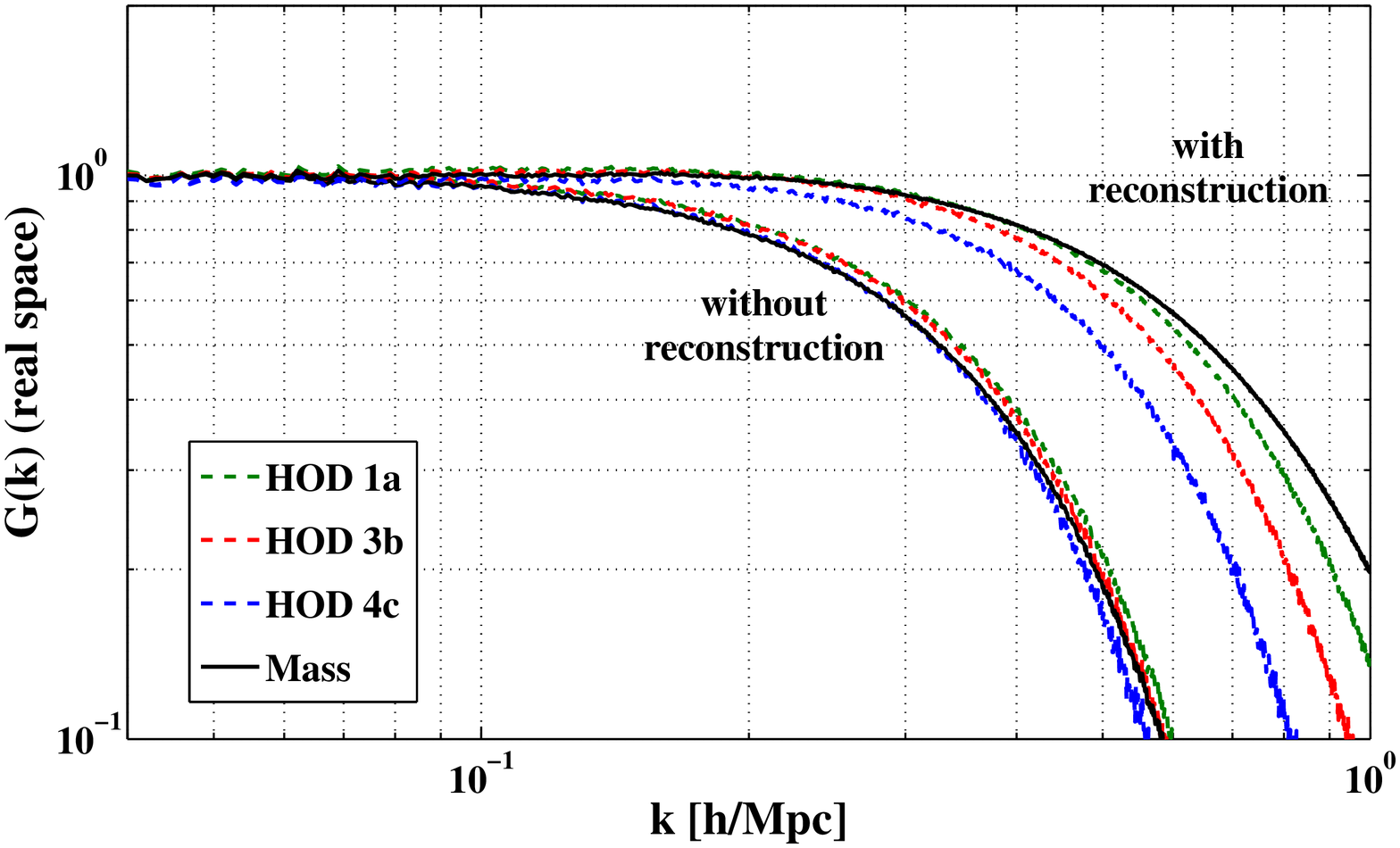}
\includegraphics[width=6.5in]{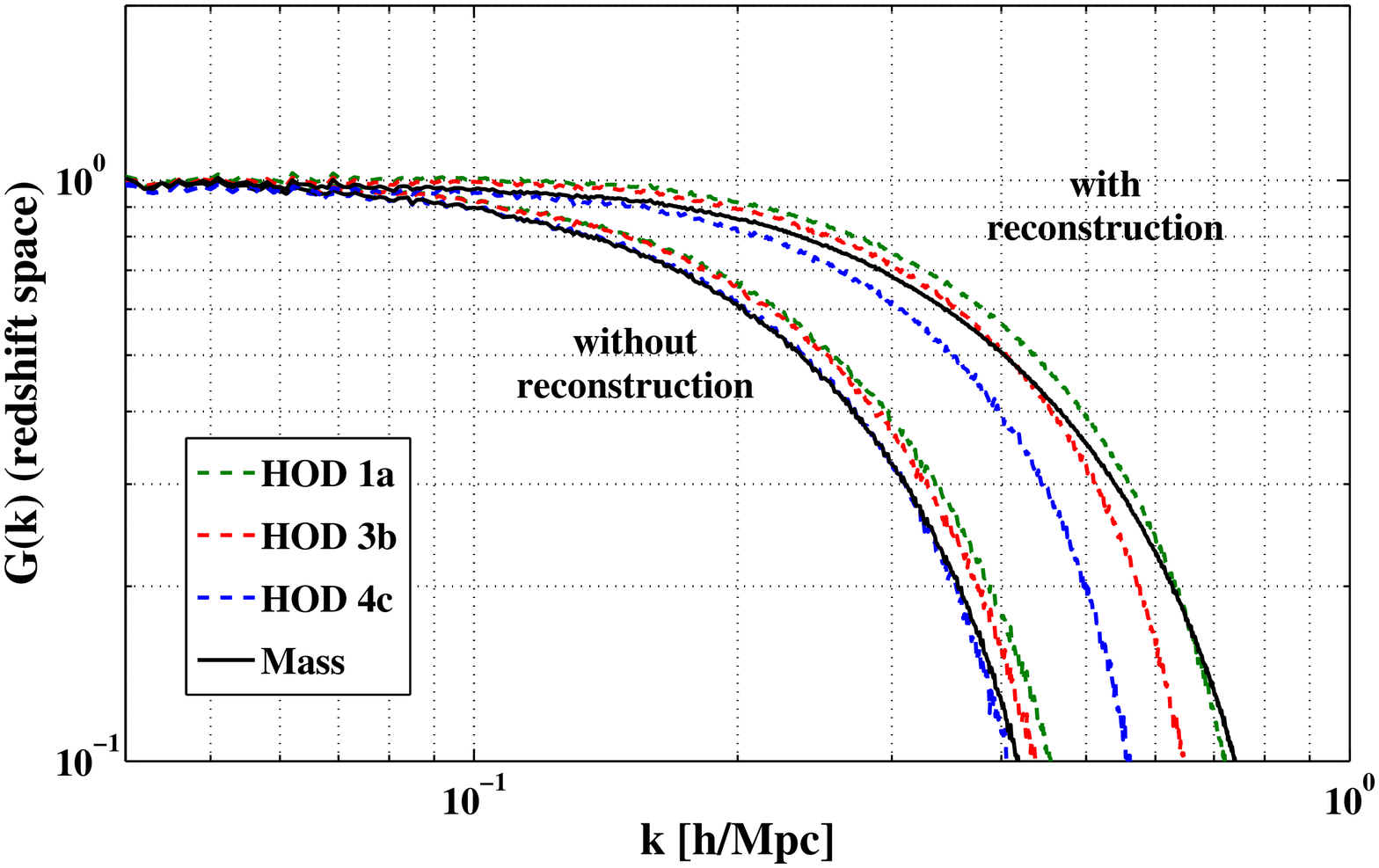}
\caption{The propagator $G(k)$ for three HOD models and includes the effects of reconstruction. The top panel shows real space and the bottom panel shows redshift space. In both panels, the lines maintaining higher correlation at smaller scales are for the cases after reconstruction. We see that in both real and redshift space, reconstruction restores information on smaller scales.}
\label{fig:CkRecon}
\end{figure}

We next investigate how reconstruction affects the density correlations in both real and redshift space. Figure~\ref{fig:CkRecon} shows the comparison of the propagator with and without reconstruction for 3 different HOD models and the mass case. We see that reconstruction restores the correlation between the initial and final ($z = 1$) density fields to larger wavenumber or smaller scales even for biased tracers with a realistic level of shot noise. In fact for $G(k) = 0.8$ the wavenumber increases by a factor of 1.5 - 2 in real space and redshift space. Reconstruction further spreads out the HOD models, which is expected as the HOD models have different shot noise levels. As reconstruction restores the correlation with the initial density fields, we expect that reconstruction will help reduce the shifts in the BAO scale and reduce the scatter on measured shifts. We discuss these effects of reconstruction in the following section. 

\section{Effect of Galaxy Bias on the Acoustic Scale Measurement}\label{sec:alpha}
\subsection{The Acoustic Scale for Biased Tracers}
\begin{figure}
\centering
\includegraphics[width=6.5in]{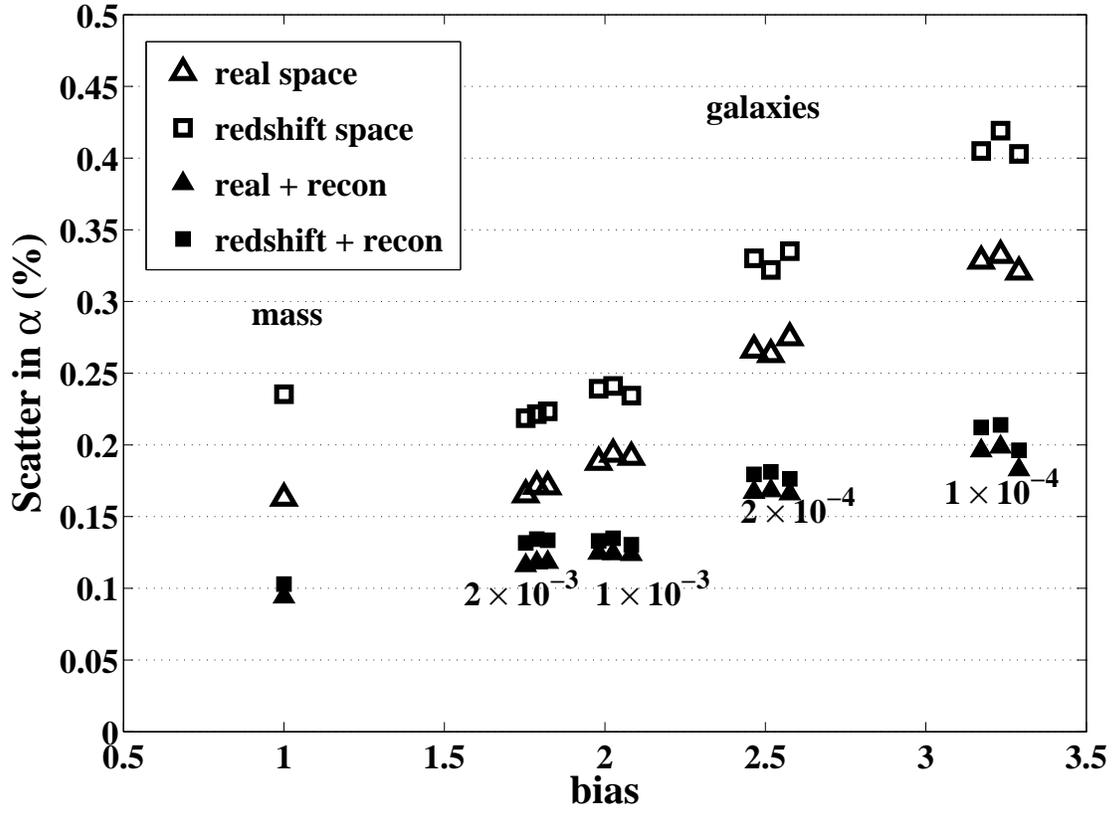}
\caption{Scatter in the $\alpha$ values as a function of bias. bias = 1 refers to the mass case. The plot also shows the effective number density ($ \rm n_{\rm eff}$) in units of $\itriMpch$ for each HOD set. From the plot, we see that reconstruction dramatically reduces the error in $\alpha$ for all HOD models in both real and redshift space. We note that, the redshift space with reconstruction errors are very close to the real space with reconstruction errors even for the highest biased case. This shows that reconstruction works well for all biased tracers.}
\label{fig:ascatter}
\end{figure}

\begin{figure}
\centering
\includegraphics[width=6.5in]{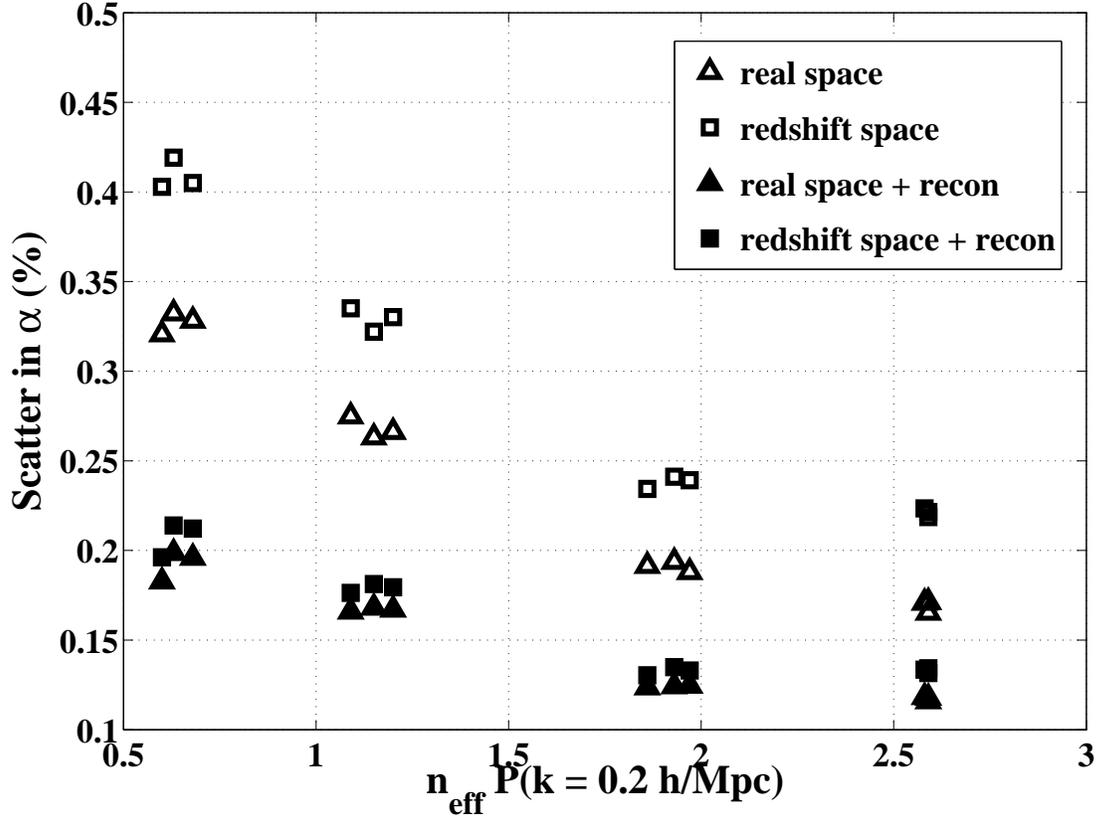}
\caption{Scatter in the $\alpha$ values as a function of $\rm n_{\rm eff}P$. From the plot, we see that the scatter in $\alpha$ reduces as we go to higher $n_{\rm eff}P$. Also, we see that reconstruction reduces the scatter in alpha consistent with the results we found and with Fig.~\ref{fig:ascatter}. }
\label{fig:neffscatter}
\end{figure}

In a measurement from a redshift survey, $\alpha$ would be the ratio of the acoustic scale in the fiducial cosmology to the measured scale. $\alpha = 1$ means that the measured acoustic scale is the same as that predicted by linear perturbation theory. In this section, we investigate the effects of realistic galaxy bias on the mean $\alpha$ and the scatter in $\alpha$ using the method described in \S~\ref{sec:Resampling}. The $\alpha$ calculated for each HOD and the mass case are given in Table~\ref{tab:alphas}. Figure~\ref{fig:ascatter} plots the scatter in $\alpha$ as a function of bias where $b = 1$ case is the mass case. Figure~\ref{fig:neffscatter} plots the scatter in $\alpha$ as a function of $\rm n_{\rm eff}P$ whose values are given in Table~\ref{tab:HOD}.

For the mass case, \cite{Seo09} used lower mass resolution but higher volume simulations (set G576 in that paper) to compute the shift in the acoustic scale and found a significant shift at $z = 1$: $\alpha_{\rm mass} - 1(\%) = 0.13 \pm 0.04$ in real space and $0.16 \pm 0.06$ in redshift space. However, the G576 simulations do not have the mass resolution to sufficiently implement our HODs. \cite{Seo09} use another set of simulations with higher mass resolution (set G1024). In the G1024 set of simulations, they find a smaller, but statistically consistent shift, with larger errors: $\alpha_{\rm mass} - 1(\%) = 0.11 \pm 0.16$ for real space and $0.002 \pm 0.23 $ for redshift space. In this paper, we will compare our results to the G1024 set of simulations from \cite{Seo09} to be consistent with the mass case. From Table~\ref{tab:alphas}, in real space, the lower biased HOD models ($b < 2.6$, sets 1,2 and 3) are consistent with the linear acoustic scale ($\alpha = 1$) within the scatter set by the sample variance of our simulations. Our highest mass HOD models (set HOD4's) shows deviation from $\alpha = 1$: $\alpha_{\rm HOD4c} - 1(\%) = 0.57 \pm 0.32$. We see a similar effect in redshift space, where HOD sets 1 and 2 are consistent with linear theory ($\alpha = 1$) and HOD sets 3 and 4 show more deviation from linear theory: $\alpha_{\rm HOD4c} - 1(\%) = 0.79 \pm 0.40$. We also note that for the biased tracers, the scatter in $\alpha$ increases with increasing bias, as we are sampling rarer halos and thus shot noise becomes more dominant in the highly biased cases. The scatter in real space ranges from $0.16\%$ to $0.26\%$ while in redshift space, it ranges from $0.22\%$ to $0.43\%$. Due to redshift distortions, we see that the scatter in redshift space is greater than in real space. 

However, once we apply reconstruction, we see that all the $\alpha$ values are consistent with unity: $\alpha_{\rm HOD4c} - 1(\%) = -0.07\pm 0.19$ in real space and $-0.11 \pm 0.20$ in redshift space. Another important difference between before and after reconstruction is the scatter in $\alpha$. We see that in all cases, the scatter in $\alpha$ is reduced by a factor of $1.5 - 2$ with reconstruction than without reconstruction in both real and redshift space which is consistent with the propagators shown in \S~\ref{sec:CkRecon}. This confirms that even for biased tracers our reconstruction technique is restoring the information and correlation between initial and final density field on the acoustic scale. This is because reconstruction corrects for the degradation of the BAO signal due to large scale bulk flows and the mass and biased tracers bulk flows only differ on small scales. From Figure~\ref{fig:ascatter}, we see that the scatter around $\alpha$ is significantly reduced in real space and in particular redshift space once we apply our simple reconstruction scheme. We also note that the redshift space with reconstruction errors are very close to the real space with reconstruction errors. We see this effect for all HOD models and hence all biased tracers. This shows us that the reconstruction scheme is bias-independent to within the scatter set by the sample variance of our simulations. We see a similar effect in Fig.~\ref{fig:neffscatter} where the scatter decreases as $\rm n_{\rm eff}P$ increases and reconstruction reduces the scatter in $\alpha$. 

\subsection{Comparing $\alpha_{\rm HOD}$ and $\alpha_{\rm mass}$}\label{sec:alphaGM}
To better understand the effects of galaxy population bias on the acoustic scale, we compare the shift in the acoustic scale ($\alpha - 1$) to the shift from the mass case. We make use of $\alpha_{\rm HOD} - 1 (\%)$ vs $\alpha_{\rm mass} - 1 (\%)$ plots. With such plots, not only can we analyze how the shifts vary between each HOD, but we can also see how they correlate with the mass case. In these plots, we plot every $\alpha - 1$ value calculated from the resampling method described above. We show one such plot of HOD 2a vs mass in Figure~\ref{fig:alphaGM}. From the figure, we see that for both real space and redshift space, there is a correlation between the shifts measured with the HOD models and with the mass case. As expected, we see that redshift space has more scatter than real space. However, we see that reconstruction not only constrains the distribution and thereby reducing the scatter around the mean measured shift, but it also increases the correlation between the HOD models and mass case.

\subsection{Difference in Acoustic Scale: $\alpha_{\rm HOD} - \alpha_{\rm mass}$}\label{sec:diffalpa}
In the previous section, we showed that there is a strong correlation between the shifts measured from the biased tracers and those measured from the mass case. This means that we can measure $\alpha_{\rm HOD} - \alpha_{\rm mass}$ more precisely than $\alpha_{\rm HOD}$ alone. This is useful because we can measure $\alpha_{\rm mass}$ more precisely using simulations with poorer mass resolution. \cite{Seo09} measured the acoustic scale at $z = 1$ in redshift space to be $\alpha_{\rm mass} - 1 (\%) = 0.158 \pm 0.061$ before reconstruction and $\alpha_{\rm mass} - 1 (\%) = 0.002 \pm 0.030$ after reconstruction using the lower resolution, larger volume simulation set (G576). Combining the sets of simulations and precisely measuring $\alpha_{\rm mass}$ and $\alpha_{\rm HOD} - \alpha_{\rm mass}$ allows us to determine $\alpha_{\rm HOD}$. 

\begin{figure}
\centering
\includegraphics[width=6.5in]{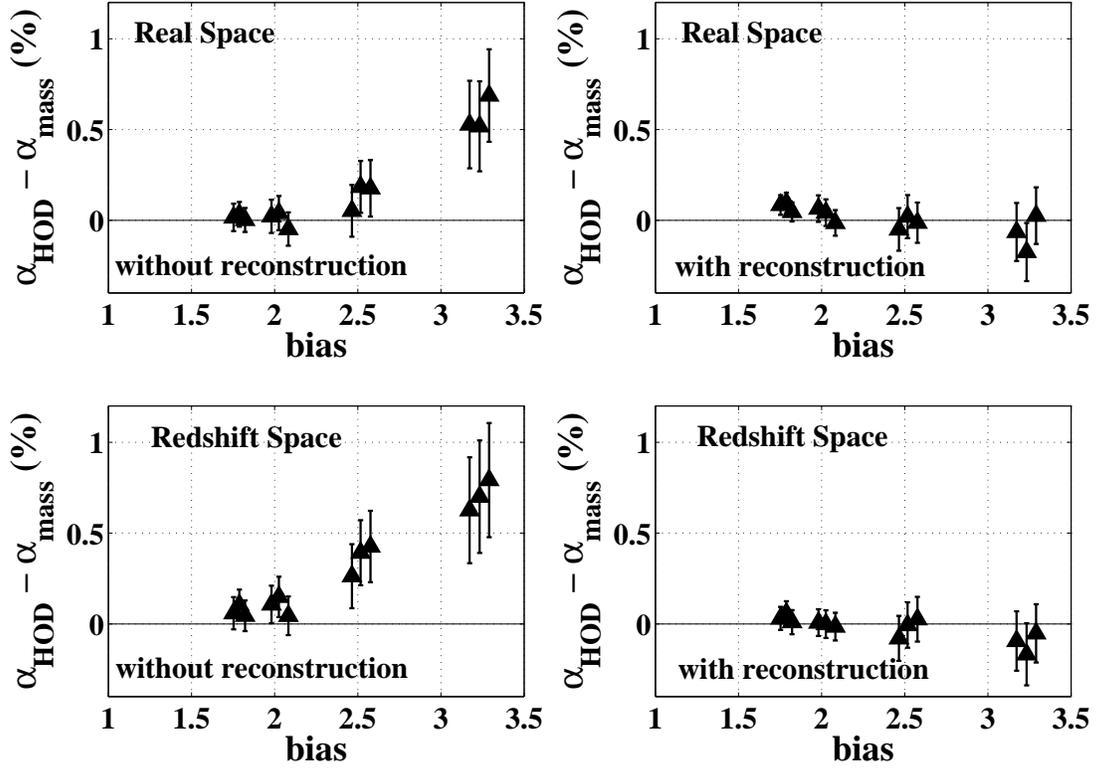}
\caption{$\alpha_{\rm HOD} - \alpha_{\rm mass}$ values for various HOD models corresponding to different biases and different number densities. We see some evidence for a shift between the most biased HODs and the mass case. However, we do not see this effect after we apply reconstruction.}
\label{fig:diffalphaGM}
\end{figure}

The $\alpha_{\rm HOD} - \alpha_{\rm mass}$ values are given in Table~\ref{tab:diffalphaGM} and shown in Figure~\ref{fig:diffalphaGM}. From the table and the figure, we see some deviation from zero difference for the high biased cases which corresponds to low number density HOD models for both real and redshift space. The difference in redshift space is slightly larger than in real space: $\alpha_{\rm HOD4c} - \alpha_{\rm mass} (\%) = 0.69 \pm 0.24$ in real space and $0.79 \pm 0.31$ in redshift space. However, the difference between the shifts from biased tracers and the mass case, even for the highest biased cases, is about $2.7\sigma$. We also note that the scatter around the difference increases with increasing bias. 

When we apply our reconstruction scheme, we see that there is no difference between the shifts measured by biased tracers and the mass case for both real and redshift space: $\alpha_{\rm HOD4c} - \alpha_{\rm mass}(\%) = -0.03 \pm 0.16$ in real space and $-0.05 \pm 0.16$ in redshift space for HOD4c. The difference between the shifts is consistent with zero with errors $\leq 0.08\%$ for HOD models 1 and 2. Equally important, we also note that reconstruction reduces the scatter around the difference in the shifts by a factor of $1.5 - 2$ depending on the HOD model. Also, the scatter around the difference in the shifts after reconstruction are the same for both real and redshift space. This is consistent with our previous result that reconstruction accounts for non linear structure formation and these flows for the biased cases and mass tracers differ only at small scales. 

\subsection{Slopes of $\alpha_{\rm HOD} - 1$ vs $\alpha_{\rm mass} - 1$ Distribution}\label{sec:slopes}
\begin{figure}
\centering
\includegraphics[width=7.0in]{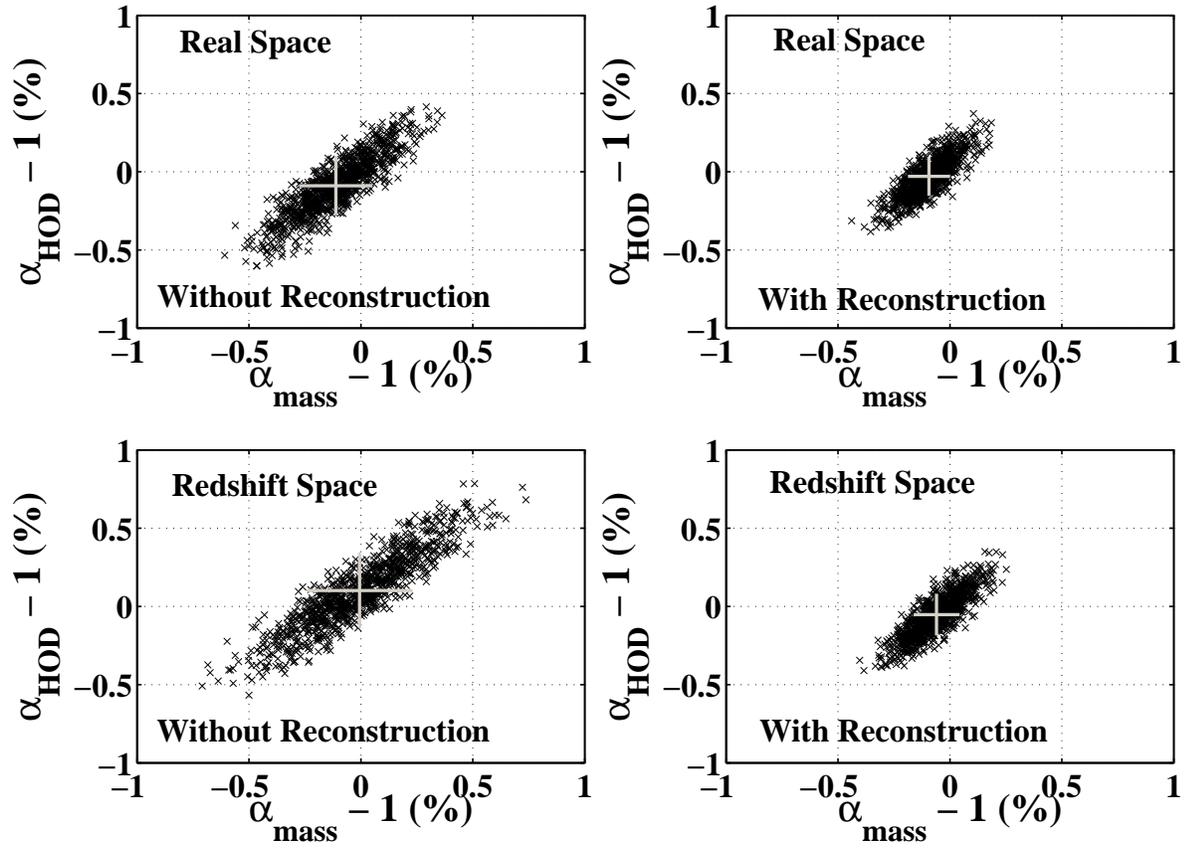}
\caption{$\alpha_{\rm HOD}$ vs $\alpha_{\rm mass}$ plot for HOD 2a for real and redshift space, with and without reconstruction. From such $\alpha-\alpha$ plots, we can compare the $\alpha$ values from the HOD models to the mass case. The crosses represent the mean $\alpha - 1$ and the scatter around the mean.}
\label{fig:alphaGM}
\end{figure}

From Figure~\ref{fig:alphaGM}, we see that the $\alpha_{\rm HOD}$ and $\alpha_{\rm mass}$ values are correlated. In this section, we investigate the slope of the $\alpha_{\rm HOD} - 1$ vs $\alpha_{\rm mass} - 1$ distribution. If the HOD models and mass cases are perfectly correlated, then we expect the underlying slope to be unity. However, there are two effects than can lead to non-unity slopes. First, if the shifts are slightly uncorrelated between the mass and HOD models, we can see a non-unity slope. Second, even if the shifts are perfectly correlated (underlying unity slope), a difference in the scatter around the mean between the mass case and the HOD models can lead to a non-unity slope. 

In Figure~\ref{fig:alphaGM}, we compute the slope by using a linear least squares fits. However, we know that this slope and the error for the slope are not correct since the linear least squares method assumes that all the scatter is around the y-axis variable, $\alpha_{\rm HOD} - 1$. We test the statistical significance of a non-unity slope by running multiple Monte-Carlo simulations with an underlying distribution with unity slope. We then compare the results of the Monte-Carlo simulations to our measurements. In detail, we use Monte Carlo simulations given by Eq.~\eqref{eq:MCalpha}, which has an underlying slope of unity to calculate the measured slope. Our model is:
\begin{align}
\alpha_{\rm mass, MC} - 1 & = \text{Gaussian(}\mu = 0,~\sigma = \sigma_{\rm mass}), \nonumber \\
\alpha_{\rm HOD, MC} - 1 & = \alpha_{\rm mass, MC} + \text{Gaussian(}\mu = 0,~\sigma = \sigma_{\rm HOD - mass})
\label{eq:MCalpha}
\end{align}
where $\sigma_{mass}$ is the scatter around $\alpha_{\rm mass}$ from Table~\ref{tab:alphas} and $\sigma_{\rm HOD - mass}$ is the scatter around $\alpha_{\rm HOD - mass}$ from Table~\ref{tab:diffalphaGM}. We use the same resampling method as used for our simulations to generate 1000 subsamples by averaging over 22 randomly selected values from 44 values for each subsample given by Eq.~\eqref{eq:MCalpha}. In order to compute the mean Monte Carlo slope and the scatter around this slope, we create 500 realizations of 1000 subsamples and use linear least squares fit. In Table~\ref{tab:GMslopes}, we give the slopes for $\alpha_{\rm HOD}$ vs $\alpha_{\rm mass}$ distribution for $\alpha_{\rm mass}$ at $z = 1$ and $z = \infty$ for four different HOD models with the other HOD models giving very similar results. We see that for both redshift space and real space, the low bias and high number density HOD models (HODs 1 \& 2) are all consistent with an underlying slope of unity. At the high bias and low number density HOD models (HODs 3 \& 4), we start to see some deviation from unity slope at the $1.5\sigma - 2.0\sigma$ level for $z = 1$ mass values. However, with reconstruction, we see that all HOD models are consistent with an underlying unity slope for both $z = 1$ and $z = \infty$ mass values within $1\sigma$.

\subsection{Fisher Matrix Scatter Estimates}\label{sec:FisherMatrix}
\cite{SE03, SE07} used Fisher matrix analyses to predict the scatter in the acoustic scale available in surveys of a given number density and bias.  These models depend on certain assumptions, such as Gaussianity of the density field up to a cut-off wavenumber, that we can check with \Nb~simulations.  Here we compare the scatter in the acoustic scale for our HOD models to those predicted by the Fisher matrix model. The Fisher matrix code takes in the number density~(Eq~\eqref{eq:neff}), $\sigma_8$~(Table~\ref{tab:HOD}), $\Sigma_{\perp}$, $\Sigma_{\parallel}$ and $\beta$. For our cosmology, at $z=1$, $\Sigma_{\perp} = 5.26$ is the rms radial displacement across the line of sight and $\Sigma_{\parallel} = 5.26$ (real space) and $9.55$ (redshift space) is the rms displacement along the line of sight~\citep{Seo09}. When we apply reconstruction, we choose $\Sigma_{\perp ,Recon} = \Sigma_{\perp}/2$ and $\Sigma_{\parallel ,Recon} = \Sigma_{\parallel}/2$. We know that the \Nb~analysis tends to overestimate the scatter in redshift spaec relative to the Fisher matrix code from~\cite{SE07} as shown by~\cite{Taka09} \&~\cite{Seo09}. This is because in our \Nb~analysis we fit a spherically averaged power spectrum, we do not optimally extract the two-dimensional information. Hence, we expect the scatter around the acoustic scale from the \Nb~simulations to be a factor of $\sqrt{1.16} = 1.08$ greater than the Fisher matrix estimates at $z = 1$.

Table~\ref{tab:alphas} provides the $\alpha - 1$ results from our simulations and the Fisher matrix estimates for the scatter are given in brackets. We see that for the low biased HOD models (HODs 1 and 2), the scatter around the mean measured $\alpha$ is very close to the Fisher matrix estimates for real space and redshift space. For the high biased HOD models (HODs 3 and 4), we see that the measured scatter is larger than the the Fisher matrix estimates by $20\% - 30\%$, which is more than expected. It is unclear if the HOD sets 3 and 4 show larger scatter due to high bias or low $n_{\it eff}P$. However, once we apply reconstruction, we have very good agreement between the measured scatter and the Fisher matrix estimates for all HOD models.

\subsection{Comparing $\alpha_{\rm real}$ and $\alpha_{\rm redshift}$}\label{sec:realred}
\begin{figure}
\centering
\includegraphics[width=7.0in]{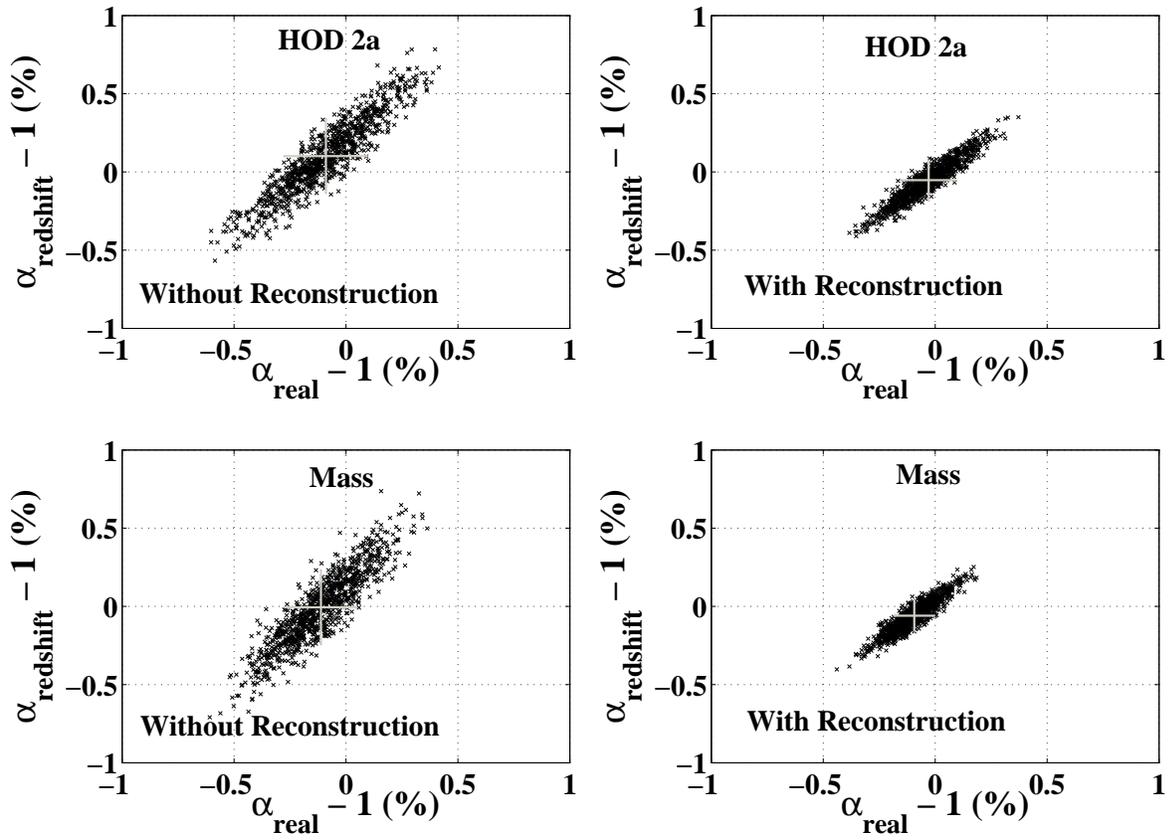}
\caption{$\alpha_{\rm real}$ vs $\alpha_{\rm redshift}$ plot for HOD2a and the mass case, with and without reconstruction. From such $\alpha-\alpha$ plots, we can study the correlation between the shifts derived from real and redshift space. The crosses represent the mean $\alpha - 1$ and the scatter around the mean $\alpha$.}
\label{fig:alphaRR}
\end{figure}

Since we measure galaxies in redshift space, we do not know the real-space positions or velocities of galaxies. In this section we see how well the shifts measured from redshift space are correlated with real space values. We use the linear least squares fit to compute the slopes of the distribution as done in \S~\ref{sec:alphaGM}~and~\ref{sec:slopes}. However, we know that not only are the shifts measured in real and redshift space correlated, but the scatter around the shifts are also correlated. In Table~\ref{tab:slopesRR}, we quote the linear least squares slopes for $\alpha_{\rm redshift}$ vs $\alpha_{\rm real}$ distribution with and without reconstruction. We expect a non-unity slope from for the distributions since we know the scatter around the redshift-space values are larger than the real space values. We see that this is true from the values in Table~\ref{tab:slopesRR}. We also see that reconstruction decreases the slope towards unity. We expect this behavior since reconstruction reduces the noise in redshift space that is uncorrelated with real space and vice-versa. Figure~\ref{fig:alphaRR} shows the $\alpha_{\rm real}$ vs $\alpha_{\rm redshift}$ plot for HOD2a and the mass case. From such plots, we can compare the shifts derived from real and redshift space. We see that reconstruction increases the correlation between real and redshift space and reduces the scatter for both sets of values. We also see that the HOD and mass cases are very similar in their real and redshift space correlation after reconstruction. The other HOD models give very similar results as shown in Table~\ref{tab:slopesRR}.

\subsection{Effects of Variations in Bias and Reconstruction Smoothing Scale}\label{sec:variations}
When we implement our reconstruction scheme outlined in \S~\ref{sec:Recon}, we use a smoothing scale of $R = 14\hMpc$. We change the smoothing scale to $R = 20\hMpc$ and rerun our reconstruction scheme. We see no change in either $\alpha$ or $\sigma_{\alpha}$ when we change the reconstruction smoothing scale. Thus, we see that reconstruction gives consistent answers over a range of smoothing scales. We also do not see any differences in $\alpha_{\rm HOD} - \alpha_{\rm mass}$ values when using the larger smoothing scale for $\alpha_{\rm HOD}$. 

\begin{figure}
\centering
\includegraphics[width=7.0in]{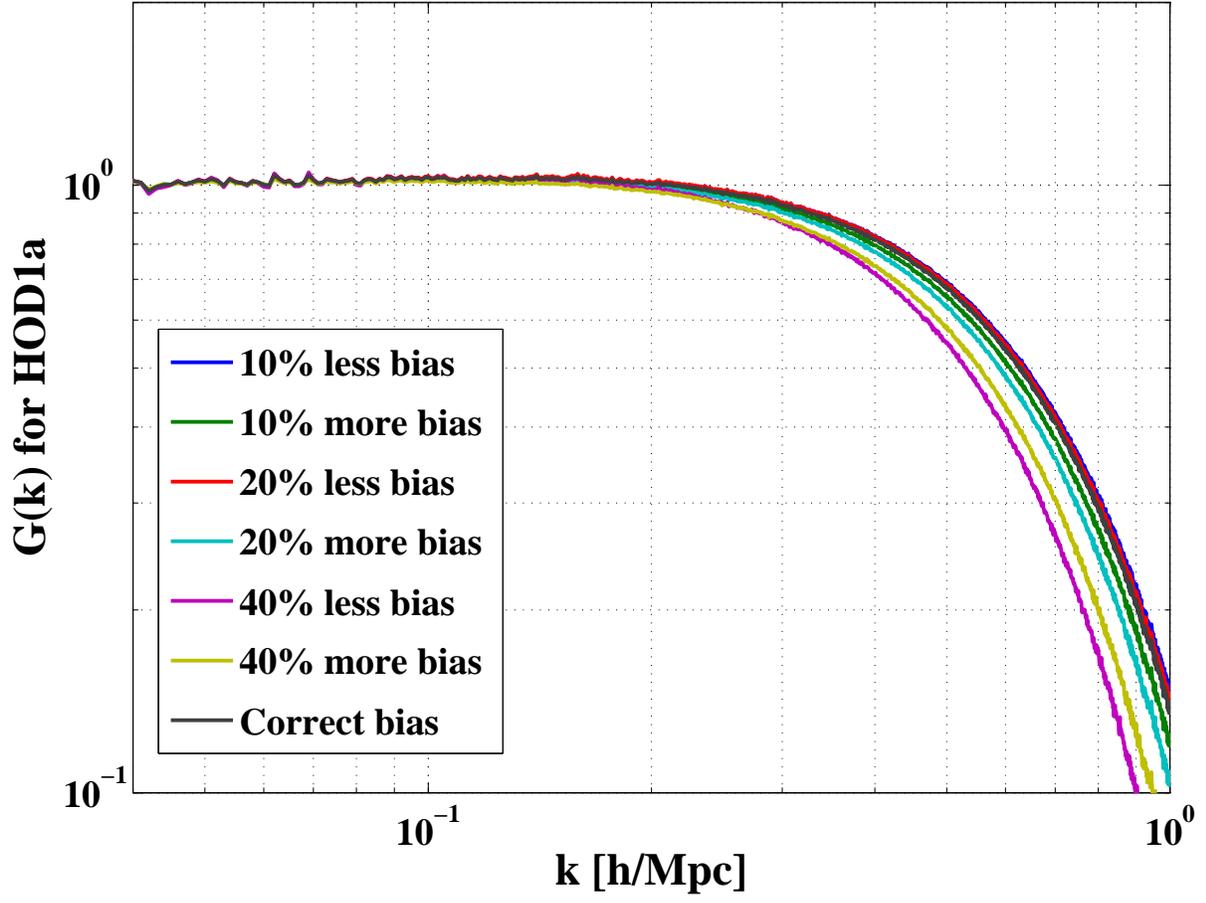}
\caption{The effect of inputing a wrong bias in the reconstruction scheme. The acoustic scale  is not affected upto a $30\%$ change in bias. We see moderate deviations from the correct bias case for the $40\%$ change in bias case.}
\label{fig:ChangingBias}
\end{figure}

To implement reconstruction, we need to estimate the value of the galaxy bias for a survey. We explore the effects of inputing a wrong bias for calculating the acoustic scale and reconstruction techniques. We use biases that are incorrect by $\pm 10\%, \pm 20\%$ and $\pm 40\%$. The results are plotted in Figure~\ref{fig:ChangingBias}. From the figure, we see that changing the bias obviously does not affect the propagator before reconstruction. After reconstruction, the propagator is very close to the actual bias case at the acoustic scale for $\pm 10\%$. We see that the $\pm40\%$ case shows moderate deviation from the correct bias case. Thus, we conclude that our reconstruction scheme is insensitive to a large range (up to $\pm 30\%$) of incorrect input bias. From future weak lensing and cluster surveys, we expect to know the amplitude of matter clustering well enough for us to know the bias to less than $10 \%$. Hence, we expect this effect to be negligible. 

\section{Conclusion}\label{sec:conc}
We use high force resolution \Nb~simulations with a total volume of $44\triGpch$ to analyze the effects of galaxy bias on the Baryon Acoustic Oscillation (BAO) scale. In particular, we apply a variety of Halo Occupation Distributions (HODs) to simulate galaxy populations and galaxy bias for upcoming dark energy surveys. We use a variety of techniques to analyze the shift in the acoustic scale for different galaxy populations and compare the derived shifts with the mass case presented in \cite{Seo09}. We extend the simple reconstruction technique introduced by \cite{ESSS07} to biased tracers and use it to remove effects of large scale gravitational flows and thereby preserving the linear theory density field. 

We use the propagator to look at the damping of the BAO signal due to non-linearities for different HOD models. We analyze our results in both real space and redshift space and apply reconstruction to both. We find that the damping is more in redshift space compared to real space. We find that galaxy bias has little effect on the propagator at BAO wavenumbers. In fact, the propagator for even the highest biased cases are close to the mass case. Reconstruction restores correlation between the initial and final density fields for both real space and redshift space. Hence, reconstruction removes most of the degradation due to non linear structure formation. We find that with reconstruction $G(k) \approx 0.9$ for $k \leq 0.3 \ihMpc$ in real space and $k \leq 0.2 \ihMpc$ in redshift space at $z = 1$. The spread in HOD models after reconstruction can be accounted by the different levels of shot noise and redshift distortions for each HOD.

We use our power spectrum fitting to measure the shift in the acoustic scale. We detect a mild shift in the acoustic scale in real space for our high bias ($b > 3$) HOD models. For the most biased HOD model, $\alpha_{\rm HOD4c} - 1(\%) = 0.57 \pm 0.32$. In redshift space, the low biased cases are consistent with no shift at $1\sigma$ while the high biased cases show a shift less than the $2\sigma$ level: $\alpha_{\rm HOD4c} - 1(\%) = 0.79 \pm 0.40$. However, once we use reconstruction, we do not detect any shift for any HOD in either real space or redshift space. Also, we see that reconstruction reduces the scatter around the measured shift by a factor of $1.5 - 2$. This confirms our results from the propagator that reconstruction undoes the degradation to the BAO signal caused by large scale bulk flows. 

In addition to the shift in the acoustic scale, we also compare the shift to the mass case. We look at the difference in the shift computed with the HOD models and the mass case and look at the scatter around this difference. We see that for the low biased HOD models, we see no difference between $\alpha_{\rm HOD}$ and $\alpha_{\rm mass}$ within $1\sigma$: $\alpha_{\rm HOD2b} - \alpha_{\rm mass}(\%) = -0.05 \pm 0.09$ in real space and $0.05 \pm 0.11$ in redshift space. The high biased HOD models show a difference at the $1.5 - 2.7 \sigma$ level: $\alpha_{\rm HOD4c} - \alpha_{\rm mass}(\%) = 0.69 \pm 0.25$ in real space and $0.79 \pm 0.31$ in redshift space. Once we apply reconstruction, the difference between the shifts are consistent with 0 within $1\sigma$. Reconstruction also reduces the errors on the difference of the shifts by a factor of $1.5 - 2$ depending on the HOD model in real and redshift space. In summary, the acoustic scale in redshift space matches that of linear theory to within $0.08 \%$ for HOD models 1 \& 2, $0.12 \%$ for HOD models 3 and $0.16 \%$ for HOD models 4.

We also test the correlation between the shifts measured with HODs and without HODs using $\alpha_{\rm HOD}$ vs $\alpha_{\rm mass}$ plots and the slopes of these distributions. From both these analyses we conclude that there is strong correlation between the shifts. We also show that reconstruction not only reduces the scatter, but also improves the correlation. With reconstruction, our results between shifts from HOD models and the mass case consistent with an underlying slope of unity within error bars. We use a similar method to analyze the correlation between shifts from real space and redshift space. We find that, as expected, reconstruction strengthens the correlation between real space and redshift space by accounting for redshift distortions. 

We use the Fisher matrix code by~\cite{SE07} to compare the measured scatter around the acoustic shift to theoretical predictions. We find that for the low biased HOD models (HODs 1 and 2), we have very good agreement between the measured scatter and the predicted scatter. For the high biased HOD models (HODs 3 and 4) however, we find that the measured scatter is larger than the predicted scatter by $20\% - 30\%$. However, after reconstruction, we have good agreement between the measured and predicted scatter for all HOD models.

Finally, we look at effects of varying the reconstruction smoothing scale and input galaxy bias. In our analysis, we have used a smoothing scale of $14 \hMpc$. We increase that to $20 \hMpc$ and we do not find any significant differences in the acoustic shift or the scatter around the shift. We considered the impact of misestimation of the bias parameter in the reconstruction method.  Varying
by $10\%, 20\%$, and $40\%$, we found mild degradations in the propagator only for the largest case. We expect future surveys to know the bias to better than $10\%$, more than adequate for reconstruction.

In conclusion, we investigate the effect of galaxy bias on the acoustic scale. We find that the effect of galaxy bias in redshift space relative to the mass case without reconstruction is about $0.1\%$ for the low bias cases ($b < 3$), growing to $0.3 \%$ in the most extreme case. With our simple reconstruction scheme, this effect is consistent with no shift with errors less than $0.08 \%$ for the low biased cases  and less than $0.16 \%$ for the high biased cases. Current surveys such as the Sloan Digital Sky Survey (SDSS) III baryon oscillations spectroscopic survey (BOSS), the WiggleZ dark energy survey, and the Hobby-Eberly telescope dark energy experiment (HETDEX) will measure the acoustic peak to about $1\%$, $2.5\%$, and $2\%$ precision respectively. While our simulations probe the effects of galaxy bias at higher redshift than current surveys, our results suggest that this effect will not be noticeable in these surveys. However, future surveys such as BigBOSS, Joint Dark Energy Mission (JDEM), and Euclid will measure the acoustic scale to a precision close to the cosmic variance limit of about 0.1\% out to $z=2$ \citep{SE07}. Our current error bars are approaching these levels, but we plan to running more high resolution simulations to lower redshifts with larger volumes in order to reduce our error bars. Future work will also look at effects of cut-sky and non-periodic boundary conditions on reconstruction.

We thank Martin White for helpful discussions. KM, DJE, JE, and XX are supported by NSF AST-0707725 and by NASA BEFS NNX07AH11G. HJS is supported by the DOE at Fermilab.

\bibliographystyle{aa}
\bibliography{GalaxyPaper}

\clearpage

\begin{deluxetable}{cccccccc}
\tablecolumns{8}
\tablewidth{0pc} 
\tablecaption{Halo Occupation Distributions (HODs) and Their Properties.}
\tablehead{
\colhead{Model} & \colhead{Total Number} & \colhead{Satellite}& \colhead{$M_{cen}$} & \colhead{$M_{sat}$} & \colhead{$\rm bias$} & \colhead{$\rm n_{eff}~(\itriMpch)$} & \colhead{$\rm {n_{\rm eff}}P(k=0.2~\ihMpc)$}\\
\colhead{} & \colhead{of Galaxies} & \colhead{Fraction ($\%$)} & \colhead{} & \colhead{} & \colhead{} & \colhead{} & \colhead{}
}
\startdata
HOD1a & $2~\times~10^6$ & 0 & 1.37 & \nodata & 1.75 & $1.14~\times~10^{-3}$ & 2.59 \\
HOD1b & $2~\times~10^6$ & 5 & 1.37 & 0.92 & 1.79 & $1.10~\times~10^{-3}$ & 2.59 \\
HOD1c & $2~\times~10^6$ & 10 & 1.37 & 0.44 & 1.82 & $1.05~\times~10^{-3}$ & 2.58 \\
HOD2a & $1~\times~10^6$ & 0 & 2.52 & \nodata & 1.98 & $6.79~\times~10^{-4}$ & 1.97 \\
HOD2b & $1~\times~10^6$ & 5 & 2.59 & 1.52 & 2.03 & $6.37~\times~10^{-4}$ & 1.93 \\
HOD2c & $1~\times~10^6$ & 10 & 2.74 & 0.74 & 2.08 & $5.80~\times~10^{-4}$ & 1.86 \\
HOD3a & $3~\times~10^5$ & 0 & 6.19 & \nodata & 2.46 & $2.67~\times~10^{-4}$ & 1.20 \\
HOD3b & $3~\times~10^5$ & 5 & 6.41 & 3.06 & 2.52 & $2.45~\times~10^{-4}$ & 1.15 \\
HOD3c & $3~\times~10^5$ & 10 & 6.70 & 1.48 & 2.58 & $2.23~\times~10^{-4}$ & 1.09 \\
HOD4a & $1~\times~10^5$ & 0 & 14.40 & \nodata & 3.17 & $9.10~\times~10^{-5}$ & 0.68 \\
HOD4b & $1~\times~10^5$ & 5 & 14.80 & 5.98 & 3.23 & $8.22~\times~10^{-5}$ & 0.63 \\
HOD4c & $1~\times~10^5$ & 10 & 15.48 & 2.88 & 3.29 & $7.51~\times~10^{-5}$ & 0.60
\enddata
\tablenotetext{a}{$M_{cen}$ is given in units of $10^{12}~h^{-1}M_{\odot}$ and $M_{sat}$ in units of $10^{14}~h^{-1}M_{\odot}$.}
\tablenotetext{b}{$\sigma_{\rm bias} = \sigma_{\rm mass}*D(z=1)/D(z=0)$ with $\sigma_{\rm mass} = 0.817$.}
\tablenotetext{c}{$\rm n_{eff}$ is given by Eq.~\eqref{eq:neff}.}
\label{tab:HOD}
\end{deluxetable}

\begin{deluxetable}{ccccc}
\tablecolumns{5}
\tablewidth{0pc} 
\tablecaption{$\alpha - 1$~(\%) values.}
\tablehead{
\colhead{Model} & \colhead{Real Space} & \colhead{Real Space} & \colhead{Redshift Space} & \colhead{Redshift Space} \\
\colhead{} & \colhead{(Theoretical)} & \colhead{with Reconstruction} & \colhead{(Theoretical)} & \colhead{with Reconstruction}
}
\startdata
mass & $-0.11 \pm 0.16~(0.15)$ & $-0.09 \pm 0.10~(0.09)$ & $-0.01 \pm 0.23~(0.21)$ & $-0.06 \pm 0.11~(0.10)$ \\
HOD1a & $-0.10 \pm 0.16~(0.16)$ & $-0.01 \pm 0.12~(0.12)$ & $-0.05 \pm 0.22~(0.22)$ & $-0.03 \pm 0.13~(0.14)$ \\
HOD1b & $-0.08 \pm 0.17~(0.16)$ & $+0.00 \pm 0.12~(0.12)$ & $+0.10 \pm 0.22~(0.22)$ & $+0.00 \pm 0.13~(0.14)$ \\
HOD1c & $-0.11 \pm 0.17~(0.17)$ & $+0.05 \pm 0.12~(0.12)$ & $+0.04 \pm 0.22~(0.22)$ & $-0.05 \pm 0.13~(0.14)$ \\
HOD2a & $-0.09 \pm 0.19~(0.17)$ & $-0.03 \pm 0.12~(0.13)$ & $+0.10 \pm 0.24~(0.23)$ & $-0.05 \pm 0.13~(0.15)$ \\
HOD2b & $-0.07 \pm 0.19~(0.18)$ & $-0.05 \pm 0.12~(0.13)$ & $+0.14 \pm 0.24~(0.24)$ & $-0.06 \pm 0.13~(0.15)$ \\
HOD2c & $-0.16 \pm 0.19~(0.18)$ & $-0.11 \pm 0.12~(0.13)$ & $+0.04 \pm 0.23~(0.24)$ & $-0.07 \pm 0.13~(0.15)$ \\
HOD3a & $-0.06 \pm 0.27~(0.20)$ & $-0.14 \pm 0.17~(0.15)$ & $+0.26 \pm 0.33~(0.27)$ & $-0.14 \pm 0.18~(0.17)$ \\
HOD3b & $+0.07 \pm 0.26~(0.20)$ & $-0.07 \pm 0.17~(0.15)$ & $+0.39 \pm 0.32~(0.27)$ & $-0.07 \pm 0.18~(0.16)$ \\
HOD3c & $+0.06 \pm 0.27~(0.20)$ & $-0.11 \pm 0.17~(0.15)$ & $+0.42 \pm 0.33~(0.27)$ & $-0.03 \pm 0.18~(0.16)$ \\
HOD4a & $+0.42 \pm 0.33~(0.25)$ & $-0.15 \pm 0.20~(0.19)$ & $+0.62 \pm 0.40~(0.33)$ & $-0.15 \pm 0.21~(0.22)$ \\
HOD4b & $+0.41 \pm 0.33~(0.25)$ & $-0.27 \pm 0.20~(0.20)$ & $+0.69 \pm 0.42~(0.33)$ & $-0.23 \pm 0.21~(0.23)$ \\
HOD4c & $+0.57 \pm 0.32~(0.26)$ & $-0.07 \pm 0.19~(0.20)$ & $+0.79 \pm 0.40~(0.34)$ & $-0.11 \pm 0.20~(0.23)$
\enddata
\label{tab:alphas}
\tablenotetext{a}{Mass values taken from~\cite{Seo09}.}
\tablenotetext{b}{The real and redshift space values given in the bracket are the theoretical Fisher matrix predictions for the scatter in $\alpha$~\citep{SE07}}
\end{deluxetable}

\begin{deluxetable}{ccccc}
\tablecolumns{5}
\tablewidth{0pc} 
\tablecaption{$\alpha_{\rm HOD} - \alpha_{\rm mass}^+$~(\%) values.}
\tablehead{
\colhead{Model} & \colhead{Real Space} & \colhead{Real Space} & \colhead{Redshift Space} & \colhead{Redshift Space} \\
\colhead{} & \colhead{} & \colhead{with Reconstruction} & \colhead{} & \colhead{with Reconstruction}
}
\startdata
HOD1a & $+0.02 \pm 0.08$ & $+0.08 \pm 0.05$ & $+0.06 \pm 0.09$ & $+0.03 \pm 0.06$ \\
HOD1b & $+0.00 \pm 0.07$ & $+0.05 \pm 0.05$ & $+0.05 \pm 0.08$ & $+0.01 \pm 0.07$ \\
HOD1c & $+0.03 \pm 0.07$ & $+0.09 \pm 0.06$ & $+0.10 \pm 0.08$ & $+0.06 \pm 0.07$ \\
HOD2a & $+0.02 \pm 0.09$ & $+0.06 \pm 0.07$ & $+0.11 \pm 0.10$ & $+0.01 \pm 0.07$ \\
HOD2b & $-0.05 \pm 0.09$ & $-0.01 \pm 0.07$ & $+0.05 \pm 0.11$ & $-0.02 \pm 0.08$ \\
HOD2c & $+0.04 \pm 0.09$ & $+0.04 \pm 0.07$ & $+0.15 \pm 0.11$ & $+0.00 \pm 0.08$ \\
HOD3a & $+0.05 \pm 0.14$ & $-0.05 \pm 0.12$ & $+0.26 \pm 0.18$ & $-0.08 \pm 0.12$ \\
HOD3b & $+0.18 \pm 0.16$ & $-0.01 \pm 0.11$ & $+0.43 \pm 0.20$ & $+0.03 \pm 0.12$ \\
HOD3c & $+0.18 \pm 0.14$ & $-0.02 \pm 0.12$ & $+0.39 \pm 0.18$ & $-0.06 \pm 0.12$ \\
HOD4a & $+0.53 \pm 0.24$ & $-0.06 \pm 0.16$ & $+0.63 \pm 0.29$ & $-0.09 \pm 0.16$ \\
HOD4b & $+0.52 \pm 0.25$ & $-0.17 \pm 0.16$ & $+0.70 \pm 0.31$ & $-0.16 \pm 0.17$ \\
HOD4c & $+0.69 \pm 0.25$ & $-0.03 \pm 0.16$ & $+0.79 \pm 0.31$ & $-0.05 \pm 0.16$
\enddata
\label{tab:diffalphaGM}
\tablenotetext{+}{Mass values taken from~\cite{Seo09}.}
\end{deluxetable}

\begin{deluxetable}{cccccc}
\tablecolumns{6}
\tablewidth{0pc} 
\tablecaption{Slopes of the $\alpha_{\rm HOD}$ vs $\alpha_{\rm mass}$ distribution.}
\tablehead{
\multicolumn{1}{c}{} & \multicolumn{2}{c}{Mass values at $z = 1$} & \colhead{~~} & \multicolumn{2}{c}{Mass values at $z = \infty$} \\ \hline
\colhead{HOD} & \colhead{Measured Slope} & \colhead{Monte Carlo Slope} & \colhead{~~} & \colhead{Measured Slope} & \colhead{Monte Carlo Slope} \\ \hline
\multicolumn{6}{c}{Real Space}
}
\startdata
HOD1b & 0.96 & $1.00 \pm 0.06$ & ~~ & 1.07 & $1.00 \pm 0.26$ \\
HOD2b & 1.04 & $0.99 \pm 0.09$ & ~~ & 1.02 & $1.00 \pm 0.31$ \\
HOD3b & 1.42 & $1.00 \pm 0.15$ & ~~ & 0.89 & $1.02 \pm 0.46$ \\
HOD4b & 1.42 & $1.00 \pm 0.24$ & ~~ & 0.71 & $0.98 \pm 0.57$ \\
\cutinhead{Real Space With Reconstruction}
HOD1b & 1.11 & $0.99 \pm 0.09$ & ~~ & 0.99 & $1.00 \pm 0.15$ \\
HOD2b & 1.07 & $1.00 \pm 0.12$ & ~~ & 0.98 & $1.00 \pm 0.17$ \\
HOD3b & 1.31 & $1.01 \pm 0.21$ & ~~ & 0.91 & $1.02 \pm 0.27$ \\
HOD4b & 1.28 & $1.01 \pm 0.27$ & ~~ & 0.98 & $0.99 \pm 0.34$ \\
\cutinhead{Redshift Space}
HOD1b & 0.88 & $1.00 \pm 0.06$ & ~~ & 1.14 & $1.03 \pm 0.35$ \\
HOD2b & 0.91 & $1.00 \pm 0.07$ & ~~ & 1.12 & $0.98 \pm 0.39$ \\
HOD3b & 1.15 & $1.00 \pm 0.12$ & ~~ & 1.02 & $0.94 \pm 0.57$ \\
HOD4b & 1.22 & $1.00 \pm 0.21$ & ~~ & 0.26 & $1.05 \pm 0.79$ \\
\cutinhead{Redshift Space With Reconstruction}
HOD1b & 1.14 & $1.00 \pm 0.10$ & ~~ & 0.96 & $1.02 \pm 0.19$ \\
HOD2b & 1.08 & $1.00 \pm 0.12$ & ~~ & 0.96 & $0.99 \pm 0.19$ \\
HOD3b & 1.31 & $1.01 \pm 0.20$ & ~~ & 0.93 & $1.00 \pm 0.28$ \\
HOD4b & 1.28 & $1.01 \pm 0.28$ & ~~ & 0.79 & $0.98 \pm 0.38$
\enddata
\label{tab:GMslopes}
\end{deluxetable}

\begin{deluxetable}{ccccc}
\tablecolumns{3}
\tablewidth{0pc} 
\tablecaption{Slopes for the $\alpha_{\rm redshift}$ vs $\alpha_{\rm real}$ distribution.}
\tablehead{
\colhead{Model} & \colhead{No Reconstruction} & \colhead{With Reconstruction}
}
\startdata
mass & 1.30 & 1.00 \\
HOD1a & 1.19 & 1.08 \\
HOD1b & 1.18 & 1.10 \\
HOD1c & 1.21 & 1.08 \\
HOD2a & 1.17 & 1.01 \\
HOD2b & 1.15 & 1.04 \\
HOD2c & 1.13 & 1.00 \\
HOD3a & 1.17 & 1.04 \\
HOD3b & 1.14 & 1.05 \\
HOD3c & 1.16 & 1.03 \\
HOD4a & 1.18 & 1.05 \\
HOD4b & 1.19 & 1.01 \\
HOD4c & 1.20 & 1.02
\enddata
\label{tab:slopesRR}
\end{deluxetable}
\end{document}